\documentclass[twocolumn,fleqn]{aastex631}

\usepackage{graphicx}
\usepackage{amsmath}
\usepackage[figure,figure*]{hypcap}
\graphicspath{{./}{figures/}}

\usepackage{rotating}
\usepackage{booktabs}

\shorttitle{Magnetic Fields of Milky Way Linear Filaments}
\shortauthors{N. K. Bhadari et al.}

\begin{document}

\title{\large The Milky Way Atlas for Linear Filaments III: Giant filaments and magnetic fields as evidence of a bubbly Galactic disk}

\correspondingauthor{Naval K. Bhadari, Ke Wang}

\author[0000-0001-8812-8460]{Naval K. Bhadari}
\altaffiliation{Boya Fellow}
\affiliation{Kavli Institute for Astronomy and Astrophysics, Peking University, 5 Yiheyuan Road, Haidian District, Beijing 100871, China}
\email{naval@pku.edu.cn}

\author[0000-0002-7237-3856]{Ke Wang}
\affiliation{Kavli Institute for Astronomy and Astrophysics, Peking University, 5 Yiheyuan Road, Haidian District, Beijing 100871, China}
\email{kwang.astro@pku.edu.cn}

\author[0000-0003-4366-6518]{Shu-ichiro Inutsuka}
\affiliation{Department of Physics, Graduate School of Science, Nagoya University, Furo-cho, Chikusa-ku, Nagoya 464-8602, Japan }

\author[0009-0003-8787-5028]{Mingke Sun}
\affiliation{Xinjiang Astronomical Observatory, Chinese Academy of Sciences, Urumqi 830011, PR China}
\affiliation{University of the Chinese Academy of Sciences, Beijing 100080, PR China}


\begin{abstract}
Linear filamentary structures are fundamental constituents of the interstellar medium and play a central role in star formation.
Their relative orientation with respect to the ambient magnetic field (B-field) provides key constraints on filament formation mechanisms. We investigate the relative orientation between Milky Way linear filaments (MWLFs) and the plane-of-sky B-field using polarization observations from the Atacama Cosmology Telescope (ACT) DR6, complemented by {\it Planck} data. Filament orientations are compared with the local B-field and the Galactic plane, while projection effects and statistical significance are quantified using Monte Carlo simulations of vector pairs in three-dimensions. We find no strong preferential alignment between MWLFs and the ambient B-field. Although the B-field is preferentially aligned with the Galactic plane with relative angles $\theta_{\rm BG}\sim0$–$25^{\circ}$, filament orientations exhibit a bimodal distribution, being either parallel or perpendicular to the plane ($\theta_{\rm FG}\sim0$–$15^{\circ}$ and $\sim75$–$90^{\circ}$). Filaments located far from the Galactic midplane ($|z|>90$ pc) preferentially show perpendicular alignment with both the plane and the B-field, whereas those near the midplane exhibit a bimodal orientation.
These results indicate that large-scale B-fields do not dominate the formation of MWLFs and instead favor a super-Alfvénic regime in which magnetic forces are dynamically subdominant, as expected for filaments associated with supernova-driven shells. Overall, our findings suggest that a face-on view of the Milky Way would resemble nearby disk galaxies such as M74, as observed in JWST images, with its disk structured by a network of supernova-driven bubbles (i.e., a bubbly disk).
\end{abstract}

\keywords{stars: formation, ISM: filaments, ISM: clouds, ISM: structure, Galaxy: structure}


\section{Introduction} \label{sec:intro}

Filamentary structures represent the fundamental architecture of the interstellar medium (ISM) and are widely recognized as the primary nurseries of star formation \citep{Molinari2010,Andre2014,Hacar2023}. Numerical simulations consistently reinforce this picture, showing that the formation of filaments and their subsequent fragmentation into star-forming cores naturally arise from the interplay of interstellar turbulence and gravity \citep[e.g.,][]{Klessen2004,Clarke2017}, with magnetic fields (B-fields) playing a pivotal role \citep[e.g.,][]{Li2014,Seifried2015,Hennebelle2019}. Despite this broad theoretical consensus, the degree to which B-fields actually shape filament formation and evolution remains actively debated, in part because most observed star-forming filaments reside within complex, multi-scale networks where interactions and overlapping structures obscure the underlying physical mechanisms \citep[e.g.,][]{Andre2017,Arzoumanian2022,Bhadari2020,Bhadari2022}. This challenge is further compounded by observational limitations: high-resolution dust-polarization measurements are typically restricted to small, targeted regions \citep[e.g.,][]{Hull2019,Sanhueza2021,Hwang2025}, while all-sky surveys such as {\it Planck} provide broad coverage but lack the angular resolution needed to probe B-field geometry within individual filaments at intermediate scales across the Galaxy \citep[e.g.,][]{Stephens2025}. As a result, establishing the true role of B-fields in filament evolution remains an open problem.

\begin{figure*}[htb!]
	\centering
	\includegraphics[width=1\textwidth]{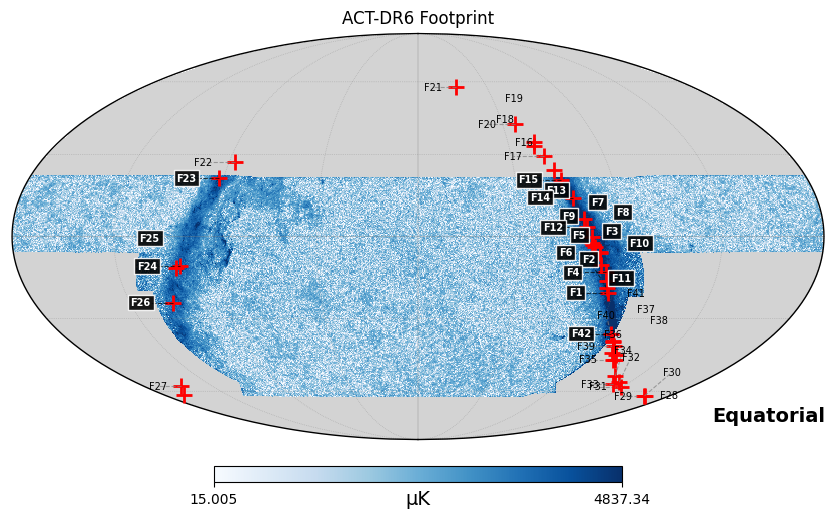}
	\caption{Galactic distribution of Milky Way linear filaments overlaid on the ACT 220 GHz intensity map in the equatorial Mollweide projection. Red crosses mark filament centers from \citet{Wang2024}, with black-box labels indicating filaments within ACT coverage (i.e., 20 MWLFs) and plain-text labels showing those outside the survey footprint.
	}
	\label{fig1:actdr6}
\end{figure*}

The relative orientation between density structures and the plane-of-sky (POS) B-field provides a powerful diagnostic of how B-fields influence molecular cloud evolution. 
Large-scale {\it Planck} observations reveal a clear trend: diffuse filamentary structures tend to align parallel to the B-field, whereas denser regions transition toward perpendicular orientations \citep{Planck2016XXXV}. This shift is often interpreted as a signature of changing cloud dynamics \citep[e.g.,][]{Soler2019,Suin2025}, and appears to occur at column densities of $N_{\rm H} \sim$ 10$^{21-22}$ cm$^{-2}$, which also corresponds to the range where the B-field strength begins to increase \citep{Crutcher2012,Chen2016}.
At the scale of individual filaments, however, both parallel and perpendicular alignments have been observed \citep[e.g.,][]{Wang2012,Palmeirim2013,Cox2016,Malinen2016,Lee2021,Butterfield2024}, a bimodal pattern also seen in Gould Belt molecular clouds \citep{Li2013} and reproduced in numerical simulations \citep{Nakamura2008,Soler2013}. These results suggest that B-fields play a dynamically significant role in filament formation. Two magnetically regulated formation mechanisms are frequently invoked: (1) magnetically channelled gravitational contraction, in which collapse occurs preferentially along field lines, producing flattened structures that fragment into filaments oriented perpendicular to the B-field \citep{Mouschovias1976,Nakamura2008}; and (2) sub-Alfvénic, anisotropic turbulence, where turbulent motions preferentially stretch gas along the field, forming filaments aligned parallel to it \citep{Stone1998,ChoVishniac2000,Vestuto2003,LiPS2008}. The relative filament–field orientation thus provides insight into the interplay between gravity and magnetically guided turbulence. 
Interpretation of these trends can be complicated by multi-scale variations in field morphology. For instance, recent Stratospheric Observatory for Infrared Astronomy (SOFIA) observations of Galactic ``bones" by \citet{Cloude2025} indicate that the B-field orientation shifts from nearly parallel on large scales (traced by {\it Planck} 850~$\mu$m at $>$10~pc) to nearly perpendicular on smaller, parsec scales (traced by SOFIA 214~$\mu$m), demonstrating that field orientation can vary significantly even along spiral arms. Bridging the gap between large-area, low-resolution and small-area, high-resolution polarization measurements is therefore essential to fully understand B-field influence on cloud and filament formation.

The Atacama Cosmology Telescope (ACT) offers a powerful new avenue for addressing this challenge. With arcminute resolution and broad sky coverage, ACT bridges the scale gap between the all-sky, low-resolution \textit{Planck} maps and high-resolution, small-area submillimeter polarimeters such as SOFIA/HAWC+ \citep{Harper2018}, JCMT/POL-2 on SCUBA-2 \citep{WardThompson2017}, and the balloon-borne BLASTPol instruments \citep{Fissel2016}. ACT's polarization maps provide both the sensitivity and angular resolution necessary to trace POS magnetic-field orientations within Galactic filaments across a large sky area \citep[e.g.,][]{Guan2021}.

In parallel, the Milky Way Atlas for Linear Filaments (hereafter, MWLF) offers an exceptional sample for this investigation. The MWLF catalog identifies 42 long, remarkably straight filaments\footnote{We refer to these structures as ``giant filaments \citep[e.g.,][]{Jackson2010,Goodman2014,Wang2015,Abreu2016,Bhadari2022},'' as their lengths (3--40 pc) and widths (0.2--2.4 pc) are substantially larger than those of typical star-forming filaments, which have characteristic widths of $\sim$0.1 pc and lengths of $\sim$1 pc \citep[e.g.,][]{Arzoumanian2011}.} traced by applying a minimum spanning tree \citep[MST\footnote{The code is available at \citet{Wang2021}, \url{https://ascl.net/2102.002}};][]{Wang2016} to \textit{Herschel} dense clumps across the Galactic plane \citep{Wang2024}. 
Their linear morphology minimizes projection effects and structural ambiguities, offering a clean laboratory for testing B-field alignment in filament evolution, a problem typically obscured in more complex filamentary networks. 
Furthermore, linear filaments have been popular targets for studying the hierarchy of massive star and cluster formation processes (e.g., \citealt{Xu2026}).
The recent ACT DR6 data release covers 20 out of the 42 MWLFs (F1-F42). Figure~\ref{fig1:actdr6} shows their Galactic distribution overlaid on the 220\,GHz intensity map from ACT DR6 \citep{Naess2025}.

The MWLF series has laid the groundwork for studying these structures: Paper I introduced the MWLF catalog and its physical properties \citep{Wang2024}, and Paper II examined the alignment between clump rotation and the parent filament \citep{Xuefang2024}. Paper I also compared filament orientations with the large-scale \textit{Planck} B-field, finding no preferred alignment.
The goal of this paper is therefore to measure, for the first time, the POS B-field orientations within MWLFs at higher angular resolution using ACT, and to assess whether these orientations reflect the influence of their ambient environment, thereby preserving signatures of their formation mechanisms.
By comparing \textit{Planck} and ACT polarization data, we also demonstrate the potential of ACT observations as a complementary resource for studying the B-fields in the ISM.
Following this introduction, Section~\ref{sec:obs} describes the ACT data and the methodology employed, Section~\ref{sec:result_dis} presents the results and discusses the implications for MWLF formation scenarios, and Section~\ref{sec:conc} summarizes the main conclusions of this work.

\begin{figure*}[htb!]
	\centering
	\includegraphics[width=1\textwidth]{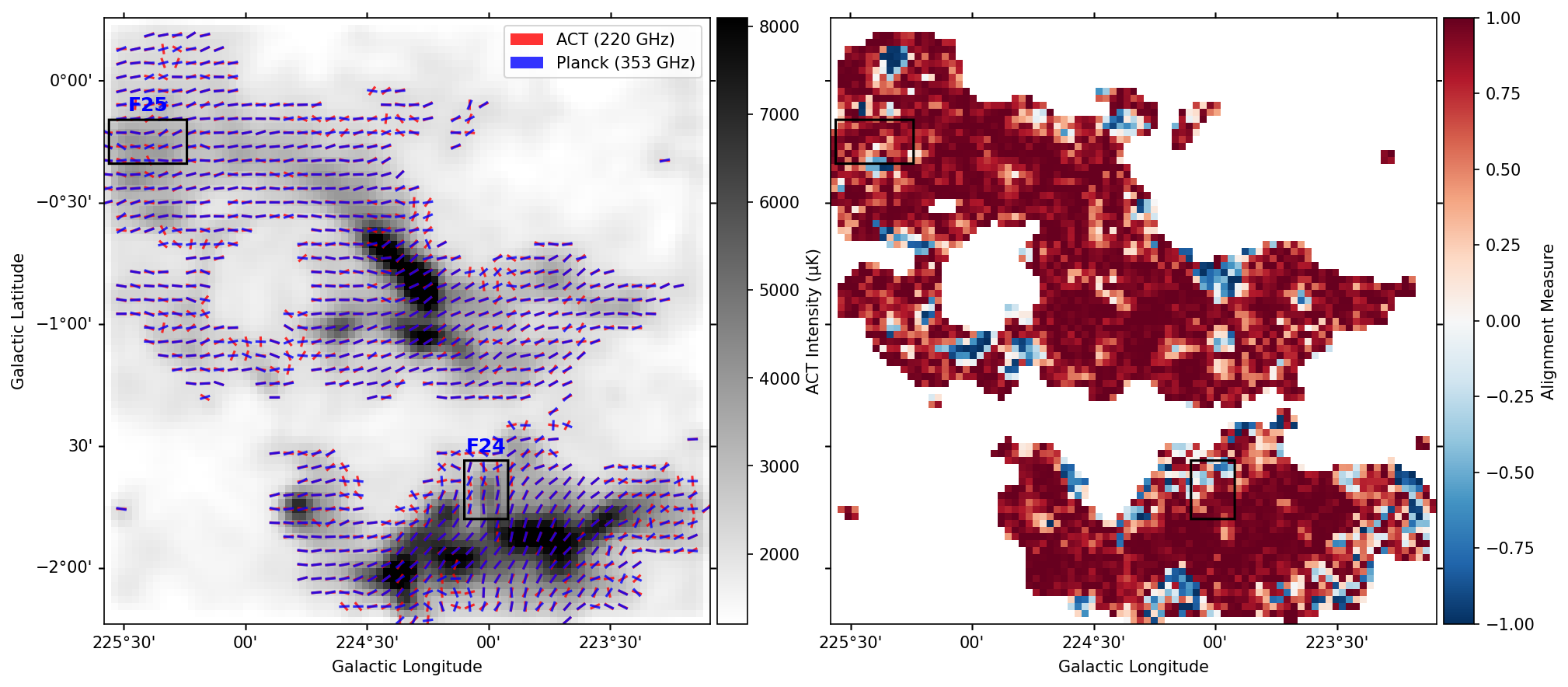}
	\caption{
Comparison between ACT and Planck-derived B-fields at 5$'$ resolution, showing B-field vector overlay on the ACT's Stokes I map (left; $I > 2000~\mu$K) and alignment measure (AM; right), where AM = $\cos(2\theta)$ and $\theta$ is the relative angle ($0^\circ$-$90^\circ$) between the two pseudovectors. AM = $+1$ indicates perfect alignment, while AM = $-1$ indicates perpendicular orientation. Red vectors represent ACT 220 GHz (ACT--{\it Planck} coadded map; see Sec.~\ref{sec:obs}), and blue vectors represent Planck 353 GHz. Boxed regions mark the footprints of the F24 and F25 MWLF targets.
This figure shows a qualitative comparison of the large-scale B-field structure in ACT and Planck. A direct comparison is provided in Figures~\ref{fig:collage1}--\ref{fig:collage3} for 20 MWLFs traced in ACT.
}
	\label{fig:Bfield_comparision}
\end{figure*}

\section{Data and Methods}
\label{sec:obs}

The ACT is a 6\,m millimeter-wave observatory located on Cerro Toco in northern Chile. It is optimized for high-resolution observations of cosmic microwave background (CMB) temperature and polarization. In this work, we make use of the recently released ACT DR6 intensity and polarization data at 220\,GHz, enabled by the upgraded Advanced ACTPol receiver, which expanded ACT’s frequency coverage to include the f220 (182--277~GHz) band \citep[see][for details]{Naess2025}. This band provides the best opportunity to trace dust polarization among the ACT channels and offers a close comparison to the \textit{Planck} 353\,GHz dust emission maps \citep[e.g.,][]{Guan2021,LuXing2024}. 
Specifically, we use the ACT--{\it Planck} coadded Stokes maps at $1'$ resolution (pixel scale $\sim30''$), constructed in the projected equatorial coordinate system. 
We adopt the ACT--{\it Planck} coadded map with day+night configuration, which provides the highest sensitivity among the available configurations, and includes point-source removal, along with the corresponding inverse-variance (uncertainty) map\footnote{ACT DR6 data are publicly available at: \url{https://lambda.gsfc.nasa.gov/product/act/act_dr6.02/}. The specific files used in this work are \texttt{act-planck\_dr6.02\_coadd\_AA\_daynight\_f220\_map\_srcfree.fits} and \texttt{act-planck\_dr6.02\_coadd\_AA\_daynight\_f220\_ivar.fits}.}.
The typical median noise levels are 230~$\mu$K ($I$) and 325~$\mu$K ($Q/U$) per native 30$''$ pixel, corresponding to approximately 115~$\mu$K\,arcmin in Stokes $I$ and 162~$\mu$K\,arcmin in Stokes $Q/U$ towards the 20 MWLFs covered in the ACT survey. For comparison, these values are 2--3 times higher than the survey averages of $\sim$40--80~$\mu$K\,arcmin in $I$ and $\sim$56--113~$\mu$K\,arcmin in polarization \citep{Naess2025}.
The elevated noise levels toward filaments may be attributed to their location near the Galactic plane.

Because our analysis is carried out in the Galactic coordinate frame, we converted the ACT Stokes parameters, originally calibrated in the equatorial frame, from ($Q, U$) to their Galactic counterparts ($Q', U'$). This transformation was applied pixel by pixel using the rotation angle between the local equatorial North and Galactic North directions \citep[e.g.,][]{Corradi1998}:

\begin{equation}
	\begin{aligned}
		\Theta(l,b)
		&= \arctan2\!\Bigl[
		\cos(l-l_{\rm ref}), \\
		&\qquad
		\cos b\,\frac{\cos b_{\rm ref}}{\sin b_{\rm ref}}
		- \sin b\,\sin(l-l_{\rm ref})
		\Bigr],
	\end{aligned}
\end{equation}

where $(l_{\rm ref}, b_{\rm ref}) = (32.9^\circ,\,62.9^\circ)$, and $(l, b)$ are the Galactic coordinates of each pixel. The corrected Stokes parameters were then obtained via
\begin{equation}
	Q' = Q \cos(2\Theta) - U \sin(2\Theta), 
	U' = Q \sin(2\Theta) + U \cos(2\Theta).
\end{equation}
From the transformed Stokes parameters $(Q',U')$, we derived the polarization angle using \mbox{$\psi_{\mathrm{pol}} = 0.5 \arctan\!2\,(U',\, Q')$} and applied a $90^\circ$ rotation to obtain the POS B-field orientation. 
The polarization intensity
	\begin{equation}
		P = \sqrt{Q'^2 + U'^2},
	\end{equation}
	and its associated uncertainty
	\begin{equation}
		\sigma_{P} = \frac{\sqrt{(Q'\,\sigma_{Q'})^2 + (U'\,\sigma_{U'})^2}}{P},
	\end{equation}
	are derived, where the per-pixel noise estimates $\sigma_{Q'} = 1/\sqrt{\mathrm{ivar}_{QQ}}$ and $\sigma_{U'} = 1/\sqrt{\mathrm{ivar}_{UU}}$ are obtained directly from the ACT DR6 inverse variance maps.
In this work, all angles are measured from Galactic north toward east in a counterclockwise sense.

To compare the ACT 220\,GHz results with those from \textit{Planck}, we additionally use the \textit{Planck} full-sky Stokes $I$, $Q$, and $U$ maps at 353\,GHz. We note that the contribution from CMB polarization at these frequencies and angular scales is negligible for ISM molecular clouds. 

\section{Results and discussion} 
\label{sec:result_dis}

In this section, we present the B-field distribution toward the MWLFs using ACT observations, validated through comparison with \textit{Planck} data, and discuss the implications for their formation and fragmentation scenarios.

\subsection{Comparison of ACT 220 GHz and Planck 353 GHz data}

To verify the consistency of our results, we first compared the ACT polarization maps with those from \textit{Planck}. The ACT maps were smoothed to the \textit{Planck} resolution of $\sim$5$'$. Figure~\ref{fig:Bfield_comparision}(a) shows the overlay of the ACT- and \textit{Planck}-derived POS magnetic-field pseudovectors on the ACT Stokes $I$ map for a larger region containing the F24 and F25 MWLFs, both located at a similar distance of about 1.2\,kpc. We quantified the alignment between the ACT and \textit{Planck} vectors using the alignment measure $AM = \cos(2\theta)$, where $\theta$ is the angle between the two vectors (ranging from $0^\circ$ to $90^\circ$), as shown in Figure~\ref{fig:Bfield_comparision}(b), with $AM=1$ indicating perfectly aligned vectors. The ACT B-field vectors exhibit a pattern consistent with that of \textit{Planck}, with $AM>0$ across the entire map, suggesting that ACT reliably traces the large-scale B-field. Similar consistency has also been reported in dedicated ACT observations of the  Central Molecular Zone (CMZ) region \citep{Guan2021,LuXing2024}.

\subsection{Filament alignment with magnetic-fields}
\label{sec:relangle_bfield}

\begin{figure}[htb!]
	\centering
	\includegraphics[width=1\linewidth]{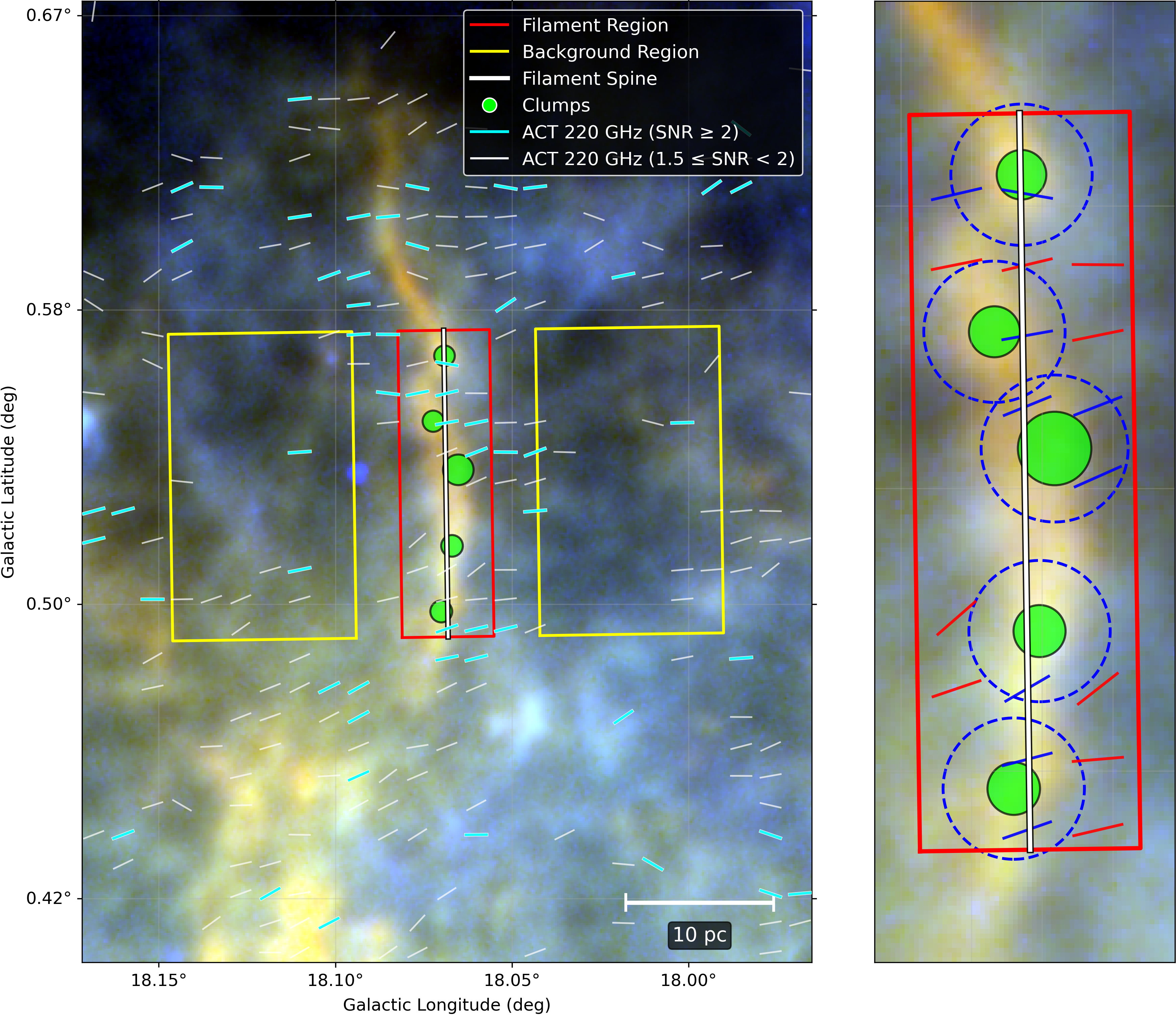}
	\caption{
			Plane-of-sky magnetic field morphology of the exemplary giant filament F4, as traced by ACT 220 GHz observations (white: $1.5 \leq P/\sigma_{\rm P} < 2$; cyan: $P/\sigma_{\rm P} \geq 2$).
			 The filament and background subregions are displayed along with the filament spine (white line) and associated clumps (lime circles). The background image is an RGB composite constructed from \textit{Herschel} 250 (Red), 160 (Green), and 70~$\mu$m (Blue) emission. The right panel presents a zoomed-in view of the filament region, where B-field vectors are color-coded: red vectors indicate positions outside clump regions, while blue vectors highlight locations within circles of 30$''$ radius (equivalent to the ACT pixel scale) around clump centers (blue dashed circles). Clumps are shown according to their reported sizes from \citet{Wang2024}. A 10 pc scale bar is shown at the bottom right of the left panel.
	}
	\label{fig:f4case}
\end{figure}

To measure the relative orientation between MWLFs and the B-field, we first determined the filament spine by performing principal component analysis (PCA) on the sky positions of clumps, weighted by the average H$_2$ column density ($N_{\rm H_2}$) of individual clumps. PCA minimizes the perpendicular distances between clumps and the fitted line, making it rotation invariant and suitable for extracting filament major axes \citep[][see Appendix B therein]{Ge2022}.
The $N_{\rm H_2}$ maps (resolution $\sim$12$''$) were obtained from a public database\footnote{http://www.astro.cardiff.ac.uk/research/ViaLactea/} and were generated using the Bayesian Point Process Mapping (PPMAP) algorithm \citep{Marsh2015,Marsh2017}, applied to Herschel 70, 160, 250, 350, and 500 $\mu$m images.
The filament angle ($\theta_{\rm F}$) is the spine orientation measured from north to east, counterclockwise.
The filament region was defined as a rectangular box  centered on the spine, with length equal to the spine and width set to three times the largest clump size, ensuring full clump coverage, which was visually verified for all targets (see Figures~\ref{fig:collage1}--\ref{fig:collage3} for comparison). The filament region, along with the filament spine, associated clumps, and overlaid B-field vectors, are illustrated for the exemplary target F4 in Figure~\ref{fig:f4case}, with similar illustrations for all 20 MWLFs covered by ACT observations shown in Figures~\ref{fig:collage1}--\ref{fig:collage3}. We also note that the presence of filaments is quite prominent in molecular line emission in comparison with the dust emission (Bhadari et al., in prep).
Stokes $(I',Q',U')$ maps within the filament region
were used to derive POS B-field angles ($\theta_{\rm B}$) following methods discussed in Sec.\,\ref{sec:obs}. Relative orientations ($\theta_{\rm FB}$) were computed as
$
|\theta_{\rm F}-\theta_{\rm B}| = \min\!\left(|\theta_{\rm F}-\theta_{\rm B}|, 
\, 180^\circ - |\theta_{\rm F}-\theta_{\rm B}| \right),
$
restricting the angle to $0-90^\circ$.  
However, the final value and uncertainty in $\theta_{\rm FB}$ were estimated using a Monte Carlo (MC) approach. For each filament, we generated 1000 realizations where Stokes $Q'$ and $U'$ at each pixel were drawn from normal distributions centered on the pixel values with standard deviations equal to the pixel-wise noise levels ($\sigma_{Q'} = 1/\sqrt{\text{ivar}_{QQ}}$, $\sigma_{U'} = 1/\sqrt{\text{ivar}_{UU}}$) obtained from the ACT uncertainty map. 
For each realization, we computed the weighted mean $\langle \theta_{\rm FB} \rangle_r$ across all filament pixels using weights $w = 1/\sigma_{\theta_B}^2$, with the B-field angle uncertainty as:
\begin{equation}
\sigma_{\theta_B} = \frac{1}{2}\,\frac{\sqrt{U'^2\,\sigma_{Q'}^2 + {Q'}^2\,\sigma_{U'}^2}}{Q'^2 + U'^2}.
\end{equation}
The final $\theta_{\rm FB}$ for the filament is the mean of these 1000 realizations, and its uncertainty is their standard deviation:
\begin{equation}
	\theta_{\rm FB} = \overline{\langle \theta_{\rm FB} \rangle_r} \pm \sigma_{\langle \theta_{\rm FB} \rangle_r}
\end{equation}
In the filament regions, $\sigma_{\theta_B}$ typically varies from $\sim\!2^\circ$ to $\sim\!11^\circ$, with a median value of $\sim\!4^\circ$.

\begin{figure*}[htb!]
	\centering
	\includegraphics[width=1\textwidth]{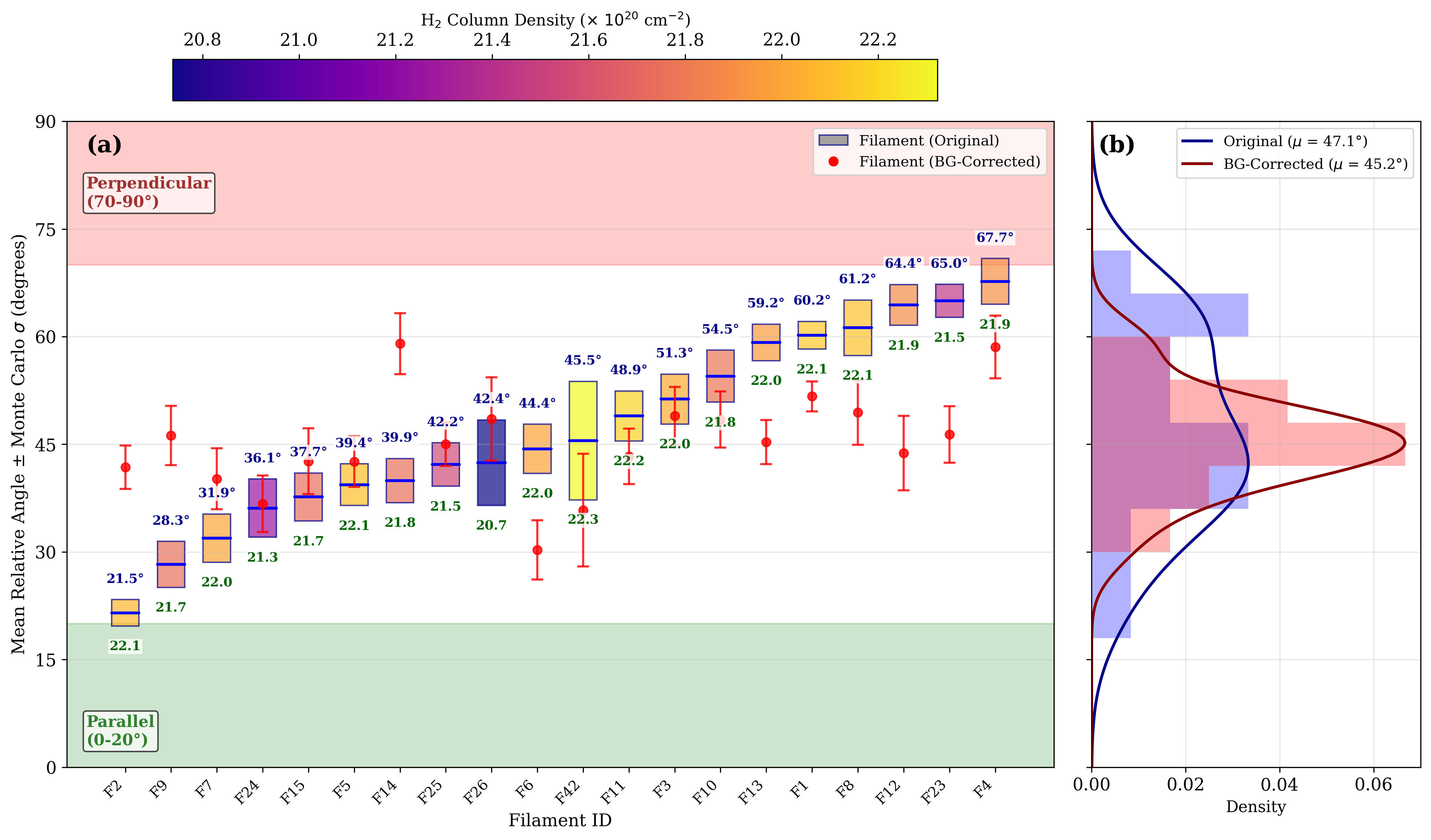}
	\caption{a) 
	Comparison of filament orientations relative to the ambient B-field. Colored boxes show Monte Carlo weighted mean relative angles $\theta_{\rm F}-\theta_{\rm B}$ (weights $= 1/\sigma_{\theta_B}^2$), with error bars representing $\pm1\sigma$ from 1000 realizations. Color-coding indicates H$_2$ column density. Red dots show background-corrected angles. Values indicate angles (top) and column density in $10^{20}$ cm$^{-2}$ (bottom). Shaded regions highlight parallel ($0^\circ$--$20^\circ$) and perpendicular ($70^\circ$--$90^\circ$) zones. (b) Histogram with Kernel Density Estimate (KDE) distributions of the relative angles $\theta_{\rm F}-\theta_{\rm B}$ for the original (blue) and background-corrected (red) measurements.
	}
	\label{fig2:relative_bangle_dist_act}
\end{figure*}

The resulting distribution of $\theta_{\rm FB}$ is shown in Figure~\ref{fig2:relative_bangle_dist_act}a, with $N_{\rm H_2}$ color-coded boxes, where the box size represents the standard deviation of the weighted mean $\theta_{\rm FB}$ obtained from 1000 MC realizations. A comparison of $N_{\rm H_2}$ with the relative angle shows no evident correlation. This is further illustrated in Figure~\ref{fig:act_planck_relanngle}a, which presents the $\theta_{\rm FB}$ distribution for all 42 MWLFs ordered by increasing $N_{\rm H_2}$.
The $\theta_{\rm FB}$ distribution shows a random orientation within an angle range of $21.5^\circ$ to $67.7^\circ$, with mean and median values of $47^\circ$ and $45^\circ$, respectively. 
Comparative plots showing the $\theta_{\rm FB}$ distribution for all 42 MWLFs using \textit{Planck} data are shown in Figures~\ref{fig:act_planck_relanngle}a and \ref{fig:act_planck_relanngle}b. 
Considering the relative angles within uncertainty, only one filament ({\bf F2}) shows a parallel alignment ($0^\circ$--$20^\circ$) in the ACT data, and {\bf F4} exhibits perpendicular alignment ($70^\circ$--$90^\circ$).
In contrast, the \textit{Planck} measurements show several cases with both parallel (i.e., {\bf F2}, F7, F24, F27, F29, F31, F32, F35, F36, and F41) and perpendicular orientations (i.e., {\bf F4}, F8, F11, F12, F13, F16, F17, F22, F23, and F33), with roughly 24\% (10/42) of filaments in each regime.


We also measured $\theta_{\rm FB}$ after correcting for foreground and background (we collectively term it as background or ``BKG") polarization following \citet[][see Sec.\,\ref{sec:fg_bg_correction} for details]{Alina2019}. 
The resulting $\theta_{\rm FB,corr}$ values are shown as red points in Figure~\ref{fig2:relative_bangle_dist_act}a, and the corresponding angle histograms of both $\theta_{\rm FB,orig}$ (standard case without BKG correction) and $\theta_{\rm FB,corr}$ (i.e., BKG corrected) are compared in Figure~\ref{fig2:relative_bangle_dist_act}b.
Unlike $\theta_{\rm FB,orig}$, the $\theta_{\rm FB,corr}$ exhibits a more pronounced peak in the $40–60^\circ$ range. 
While both distributions show a modest enhancement in this interval, only $\theta_{\rm FB,orig}$ shows good agreement with the tested MC models based on the Kolmogorov--Smirnov (KS) test. The details are discussed in Sec.\,\ref{sec:galactic_dynamics}.

\subsection{Effect of  Galactic dynamics}
\label{sec:galactic_dynamics}

\begin{figure*}[htb!]
	\centering
	\includegraphics[width=1\textwidth]{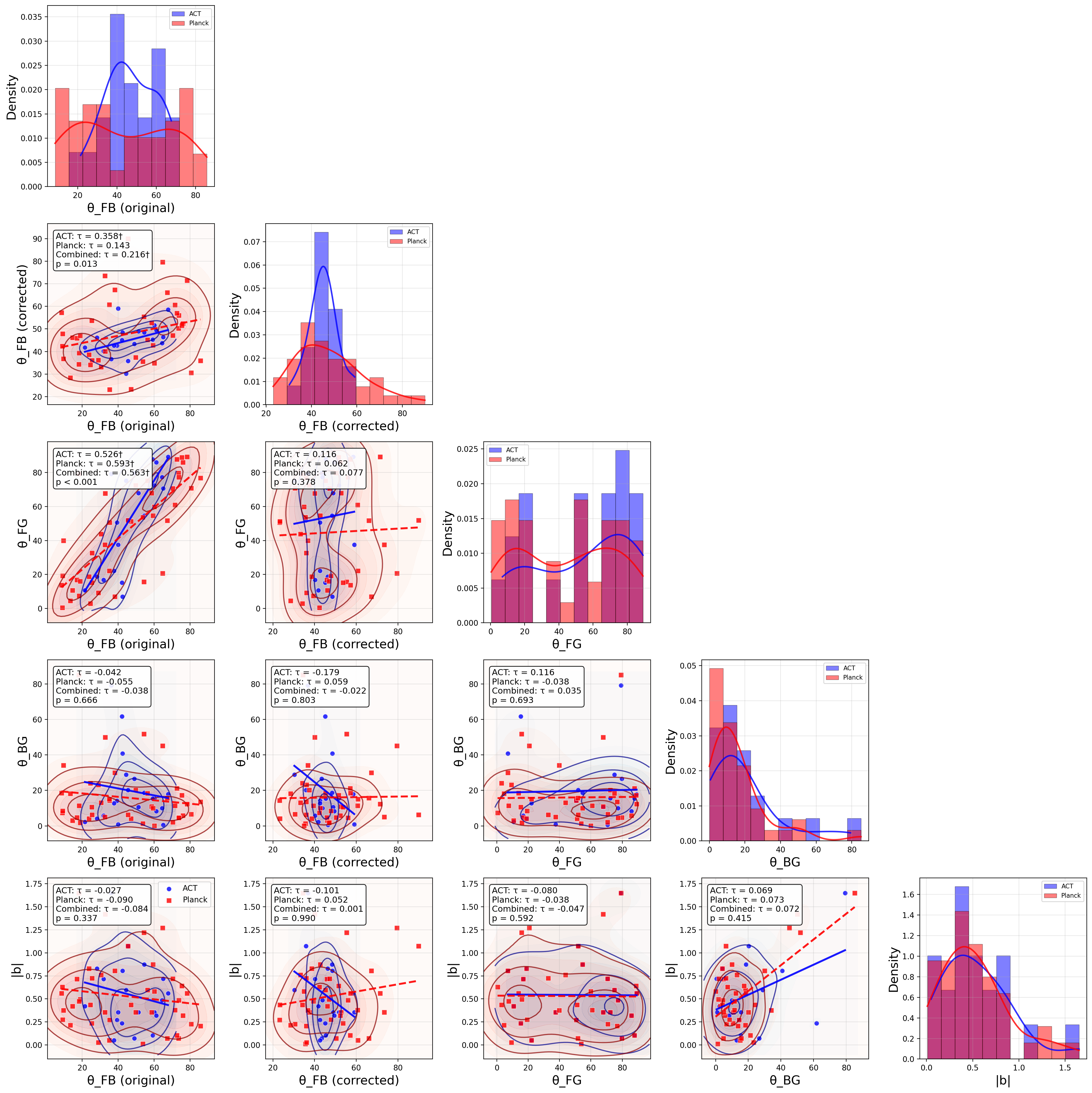}
	\caption{
		Scatter plot matrix (corner plot) showing correlations among $\theta_{\rm FB,orig}$, $\theta_{\rm FB,corr}$, $\theta_{\rm FG}$, $\theta_{\rm BG}$, and Galactic latitude ($|b|$) for ACT (blue, 20 MWLFs) and \textit{Planck} (red, 42 MWLFs). Diagonal panels show histograms with KDE. Lower panels display scatter points, 2D density contours, and best-fit linear trends (solid for ACT, dashed for \textit{Planck}). Kendall's $\tau$ (rank correlation coefficient) is reported for each dataset, with $\dagger$ denoting $p < 0.05$. The $p$-value for combined data is also shown.
	}
	\label{fig:correlation_analysis}
\end{figure*}

To assess whether the filament–B-field orientations are influenced by Galactic dynamics, such as their spatial distribution within the Galaxy or the large-scale Galactic B-field, we also examined the orientations of both the filaments and the local B-field with respect to the Galactic plane. 
Figure~\ref{fig:correlation_analysis} shows the set of correlation plots for all  relative-angle parameters studied here. These include the filament–B-field angle without BKG correction ($\theta_{\rm FB,orig}$), the BKG-corrected filament–B-field angle ($\theta_{\rm FB,corr}$), filament–Galactic-plane angle ($\theta_{\rm FG}$), the B-field–Galactic-plane angle ($\theta_{\rm BG}$), and Galactic latitude ($|b|$).
The notable features are a bimodal distribution of $\theta_{\rm FG}$ and parallel alignment between B-field and Galactic-plane. 
Correlations were quantified using Kendall's $\tau$ rank correlation coefficient. A moderate monotonic correlation (Kendall's $\tau \sim 0.56$) is seen between $\theta_{\rm FB,orig}$ and $\theta_{\rm FG}$ in both ACT and \textit{Planck}. The positive correlation arises from the combination of the weakly bimodal (effectively random) distribution of $\theta_{\rm FB,orig}$ and the clearly bimodal distribution of $\theta_{\rm FG}$. This correlation is expected, as the B-field is preferentially aligned with the Galactic plane, causing filament orientations with respect to both to exhibit similar trends. 
We find no significant correlation between $\theta_{\rm BG}$ and $|b|$ (Kendall's $\tau = 0.07$, $p = 0.415$), indicating that the orientation of the magnetic field relative to the Galactic plane does not depend systematically on Galactic latitude. The apparent trend suggested by the Pearson's correlation test, however, shows a moderate correlation ($r \sim 0.54$), demanding verification with a larger sample.

\begin{figure*}[htb!]
	\centering
	\includegraphics[width=1\textwidth]{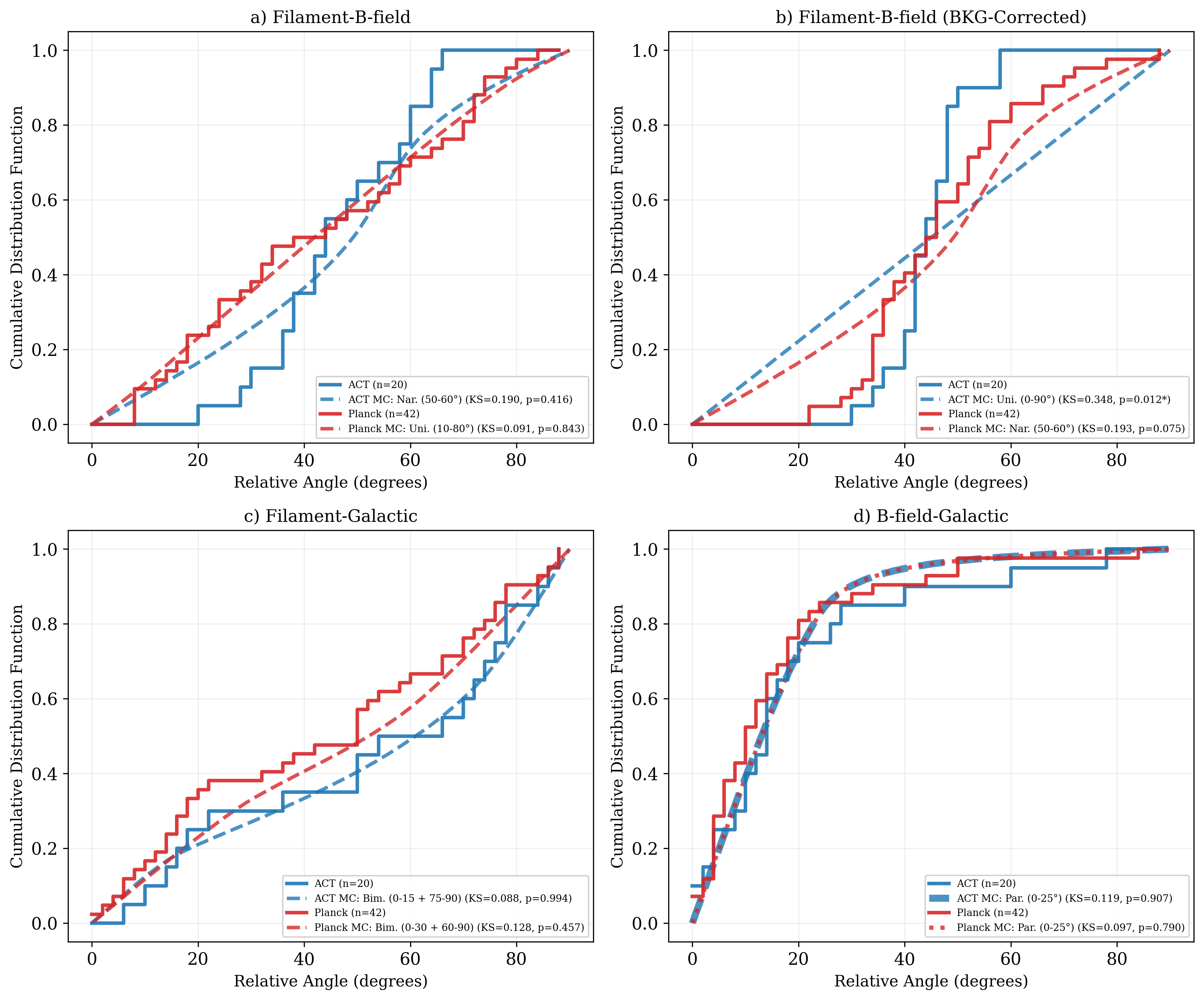}
	\caption{
			Cumulative distribution functions of the relative angles $\theta_{\rm FB,orig}$, 
			$\theta_{\rm FB,corr}$, $\theta_{\rm FG}$, and $\theta_{\rm BG}$ derived from ACT and \textit{Planck} 
			observations. The dashed curves show the best among tested MC simulation models of 
			projected relative angles for the corresponding intrinsic orientation ranges. 
			KS test statistics for each data--simulation comparison are labelled.
	}
	\label{fig:cdf_comparison}
\end{figure*}

Figure~\ref{fig:cdf_comparison} shows the cumulative distribution functions (CDFs) of $\theta_{\rm FB,orig}$, $\theta_{\rm FB,corr}$, $\theta_{\rm FG}$,  and $\theta_{\rm BG}$.
To assess the impact of geometric projection effects, we performed MC simulations of one million randomly oriented three-dimensional (3D) unit-length vector pairs. From these, we selected subsets whose intrinsic 3D angle fell within specified ranges of interest. We then projected these selected vectors onto the plane of the sky to obtain 2D angles for comparison with observations. We tested a wide range of models, including parallel cases ($0^\circ$--$15^\circ$, $0^\circ$--$20^\circ$, $0^\circ$--$25^\circ$, $0^\circ$--$30^\circ$), perpendicular cases ($60^\circ$--$90^\circ$, $65^\circ$--$90^\circ$, $70^\circ$--$90^\circ$, $75^\circ$--$90^\circ$), intermediate narrow cases ($30^\circ$--$50^\circ$, $40^\circ$--$50^\circ$, $50^\circ$--$60^\circ$, $50^\circ$--$70^\circ$), bimodal cases (($0^\circ$--$15^\circ$ + $75^\circ$--$90^\circ$), ($0^\circ$--$20^\circ$ + $70^\circ$--$90^\circ$), ($0^\circ$--$25^\circ$ + $65^\circ$--$90^\circ$), ($0^\circ$--$30^\circ$ + $60^\circ$--$90^\circ$)), and random uniform cases ($0^\circ$--$90^\circ$, $10^\circ$--$80^\circ$, $20^\circ$--$70^\circ$). For each model, we performed a KS test comparing the simulated 2D angle distribution with the observed data to quantify the similarity between the observed and simulated distributions.  The KS statistic ($D$-value) measures the maximum deviation between the observed and model CDFs, and we identify the best-fitting model as the one that minimizes this deviation. In this context, the KS test is used as a comparative goodness-of-fit metric rather than a formal null hypothesis test. Accordingly, the associated p-values are only used as an indication of whether a model can be rejected at a given significance level (i.e., $<0.05$). The best-fitting MC CDFs, along with their corresponding p-values, are overplotted in Figure~\ref{fig:cdf_comparison}. 

ACT shows a modest peak in $\theta_{\rm FB,orig}$ at 50--60$^\circ$ ($D=0.19$, $p = 0.42$), likely reflecting the smaller number of filaments covered by ACT (20) compared to \textit{Planck} (42), for which the distribution is consistent with being random over 10--80$^\circ$ ($D=0.09$, $p = 0.84$).
For the $\theta_{\rm FB,corr}$ distribution, no tested MC model provides a satisfactory match to the ACT data. In contrast, the closest match for {\it Planck} is found for the 50--60$^\circ$ model, which yields the lowest KS statistic ($D=0.19$, $p = 0.08$) among the tested models.
This indicates that the observed $\theta_{\rm FB,corr}$ distribution is not fully reproduced by randomly oriented 3D vectors projected to 2D. This discrepancy may arise from non-uniform background polarization, local environmental effects, or other factors not captured in the simplified MC model, and warrants further investigation in future studies.
Hereafter, $\theta_{\rm FB}$ refers to $\theta_{\rm FB,orig}$ unless stated otherwise. The derived physical parameters of 20 MWLFs traced by ACT are summarized in Table~\ref{tab:act_filament_results}.

\begin{figure*}[htb!]
	\centering
	\includegraphics[width=1\textwidth]{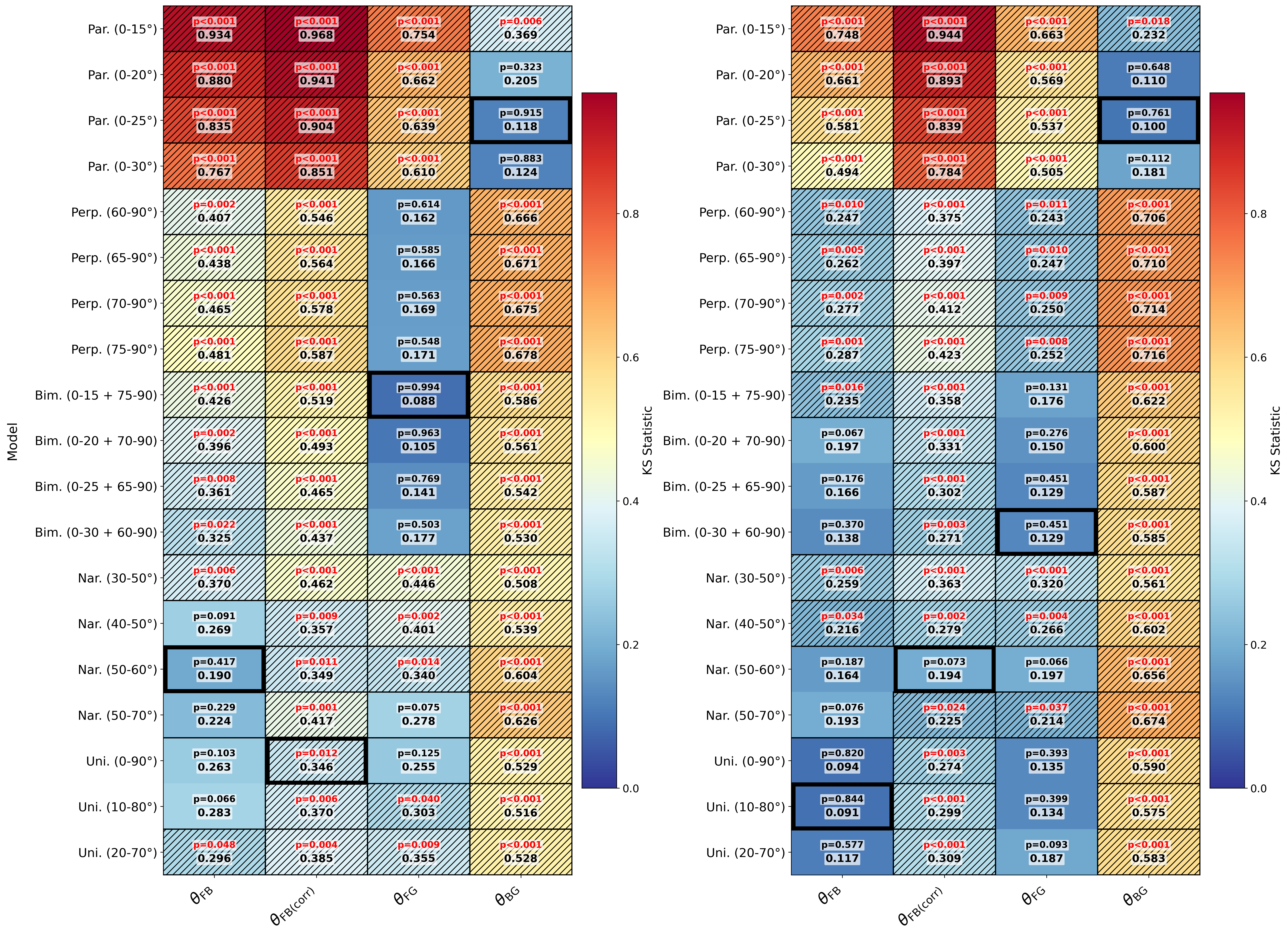}
	\caption{
			Kolmogorov-Smirnov (KS) statistics comparing the observed angle distributions with simulated Monte Carlo models for ACT (left) and {\it Planck} (right). Colors indicate the magnitude of the KS statistic. Hatched cells denote models rejected at $p < 0.05$, and black boxes highlight the best-fitting models corresponding to the minimum KS values. KS statistics and their associated $p$-values are labeled in each cell.
	}
	\label{fig:ks_heatmap}
\end{figure*}

A detailed comparison between all tested models is presented as a KS statistic heat map in Figure~\ref{fig:ks_heatmap}, with the best-fitting models highlighted by black boxes. The overall results show that, under the null hypothesis test, the parallel models ($0^\circ$--$20^\circ$, $0^\circ$--$25^\circ$, and $0^\circ$--$30^\circ$) cannot be rejected for $\theta_{\rm BG}$, while the bimodal models (($0^\circ$--$15^\circ$ + $75^\circ$--$90^\circ$), ($0^\circ$--$20^\circ$ + $70^\circ$--$90^\circ$), ($0^\circ$--$25^\circ$ + $65^\circ$--$90^\circ$), and ($0^\circ$--$30^\circ$ + $60^\circ$--$90^\circ$)) cannot be rejected for $\theta_{\rm FG}$.
A bimodal distribution of $\theta_{\rm FG}$ is also observed in the the Structure, Excitation, and Dynamics of the Inner Galactic InterStellar Medium \citep[SEDIGISM;][]{Ge2023}, and the the Milky Way Imaging Scroll Painting project \citep[MWISP;][]{Ge2024} filaments, consistent with scenarios in which Galactic dynamics and supernova-driven feedback produce preferred parallel and perpendicular alignments \citep{Joung2006,Inutsuka2015,Zucker2019}. However, this trend appears weaker in the Bolocam Galactic Plane
Survey \citep[BGPS;][]{Wang2016} and the APEX Telescope Large Area Survey of the Galaxy \citep[ATLASGAL;][]{Ge2022} samples, likely due to differences in filament tracers.
In contrast, the B-field is preferentially aligned with the Galactic plane, indicating the dominant influence of the large-scale Galactic B-field \citep[e.g.,][]{Cloude2025}.

To examine whether the filament population behaves differently as a function of vertical distance from the Galactic midplane ($|z|$), we analyze the distributions of $\theta_{\rm FB}$ and $\theta_{\rm FG}$ as functions of $|z|$. We performed a two-sample Kolmogorov--Smirnov test scan over $|z|$ thresholds from 20 to 250 pc in steps of 5 pc, comparing filaments below and above each threshold. The minimum $p$-value, corresponding to the strongest separation between the two subsamples (i.e., distinct behavior below vs. above the threshold), occurs at $|z| \sim 90$ pc (left panel of Figure~\ref{fig:relangle_height}), supporting this value as the natural division point, although the significance does not reach the formal $p<0.05$ threshold to confidently consider these as two distinct populations. 
A threshold at $|z| = 90$ pc likely separates two filament populations: filaments near the Galactic midplane exhibit both parallel and perpendicular alignments, whereas filaments farther from the midplane are preferentially perpendicular to the plane (right panel of Figure~\ref{fig:relangle_height}).

\begin{figure*}[htb!]
	\centering
	\includegraphics[width=1\linewidth]{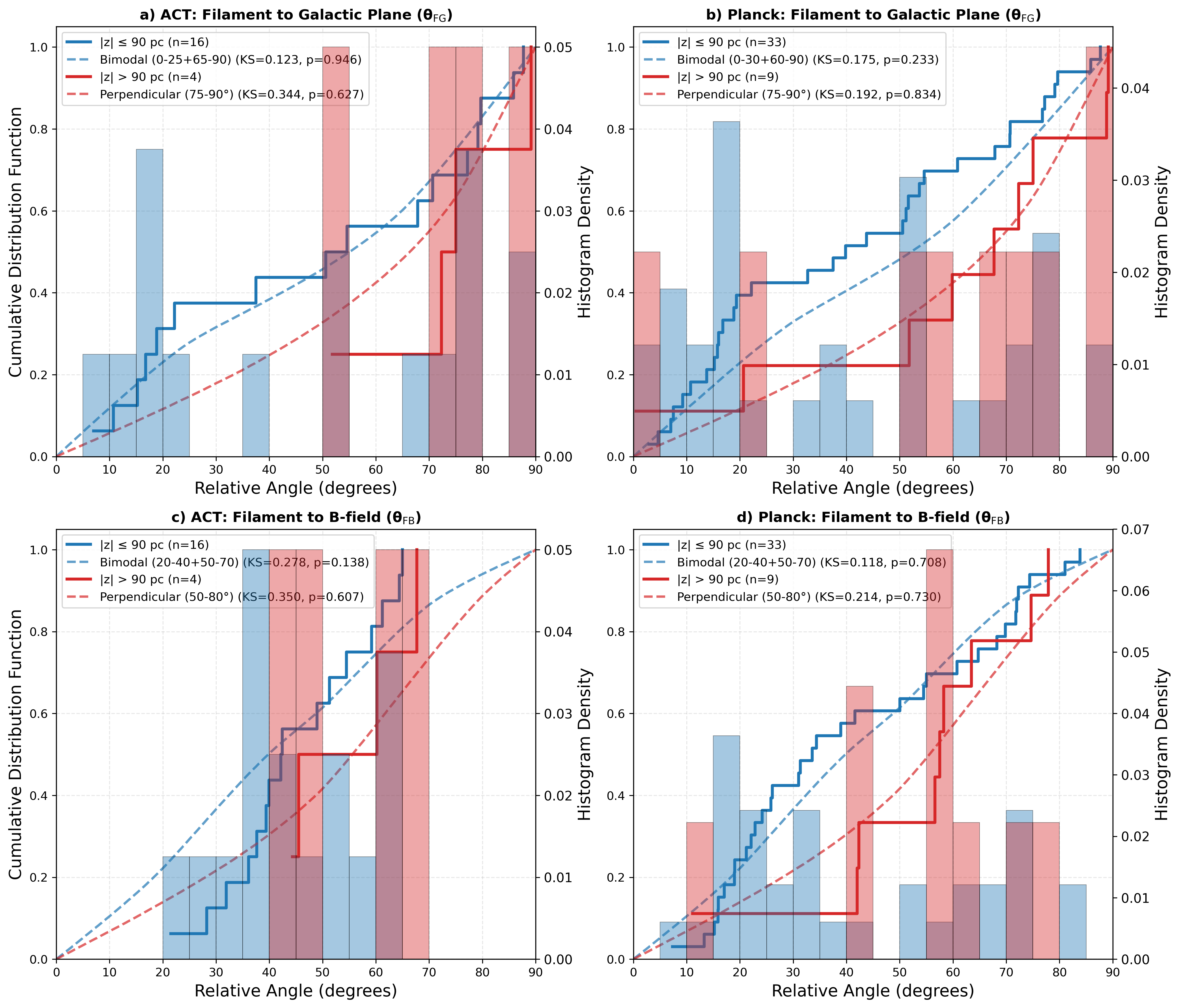}
	\caption{
Cumulative distribution functions of $\theta_{\rm FG}$ (top) and $\theta_{\rm FB}$ (bottom) for filaments with $|z| \leq 90$ pc (blue) and $|z| > 90$ pc (red). Histograms (shaded) show the observed distributions. Dashed curves show the Monte Carlo models for bimodal and perpendicular cases (KS statistic and $p$-value in legend). The left column shows results for ACT (20 MWLFs), while the right column shows results for \textit{Planck} (42 MWLFs).}
	\label{fig:relangle_histogram}
\end{figure*}

Figure~\ref{fig:relangle_histogram} presents the CDFs and histogram distributions of $\theta_{\rm FG}$ (top panels) and $\theta_{\rm FB}$ (bottom panels) for MWLFs observed with ACT (20 filaments) and {\it Planck} (42 filaments). For filaments located near the Galactic midplane ($|z| \leq 90$ pc), $\theta_{\rm FG}$ exhibits a bimodal distribution, consistent with the 0--30$^\circ$ + 60--90$^\circ$ model, whereas filaments at larger vertical distances ($|z| > 90$ pc) preferentially show perpendicular alignment, consistent with the 75--90$^\circ$ model. A similar trend is observed in the $\theta_{\rm FB}$ distributions, with midplane filaments consistent with a bimodal 20--40$^\circ$ + 50--70$^\circ$ model and higher-$|z|$ filaments consistent with a 50--80$^\circ$ model. 
For comparison, we also overplot MC simulations for representative bimodal and perpendicular cases. 
These are not intended to represent the unique best models among all the tested cases listed earlier, but they demonstrate the existence of bimodal (0--30$^\circ$ + 60--90$^\circ$ for $\theta_{\rm FG}$ and 20--40$^\circ$ + 50--70$^\circ$ for $\theta_{\rm FB}$) and perpendicular models (75--90$^\circ$ for $\theta_{\rm FG}$ and 50--80$^\circ$ for $\theta_{\rm FB}$) that cannot be rejected by null hypothesis testing.

The behavior shown in Figure~\ref{fig:relangle_histogram} is consistent with filament formation in supernova-driven shells/bubbles \citep{Inutsuka2015}, where shells can be approximated as short cylindrical structures with filaments forming along their walls, i.e., preferentially perpendicular to the Galactic plane. Filaments near the Galactic midplane therefore comprise two populations: those associated with bubble walls and those confined to the midplane, tracing dense gas in the Galactic disk \citep[e.g., Galactic bones;][]{Goodman2014,Cloude2025}. This naturally explains the observed bimodal distribution of $\theta_{\rm FG}$.
In general, filaments formed by supernova shock waves are expected to be oriented perpendicular to the B-field in three-dimensional space, consistent with the preferential perpendicular alignment observed for filaments at $|z| > 90$ pc. By contrast, near the midplane ($|z| \leq 90$ pc), where the B-field tends to align with the Galactic plane, filaments are also more likely to be parallel to the B-field, giving rise to the observed bimodality.
Overall, our results point to a broader picture of the Milky Way as a disk structured by supernova-driven bubbles (i.e., a ``bubbly'' disk), resembling nearby face-on spiral galaxies such as M74 as observed by the James Webb Space Telescope (JWST).\footnote{\url{https://esawebb.org/images/potm2208a/}}
This interpretation is further supported by the Galactic supernova rate of approximately one event every $\sim$100 yr, which implies the presence of a few thousand supernova remnants in the Milky Way. For comparison, \citet{Watkins2023} classified 1694 structures as bubbles with radii between 6 and 552 pc in the galaxy M74.
Additionally, a handful of Milky Way studies have directly linked molecular clouds and giant filaments to H\,{\sc i} supershells. \citet{Dawson2015} found that the GMC G288.5+1.5 formed at the interface of two colliding H\,{\sc i} supershells, while \citet{Clarke2023} showed that the giant molecular filament G214.5$-$1.8 resides on the wall of an H\,{\sc i} superbubble. Together, these studies reinforce the bubbly, feedback-driven picture of the Galactic disk.

\subsection{Filament formation and fragmentation scenario}
\label{mwlf:formation_scenario}

Some cloud-formation models suggest that large-scale Galactic B-fields have little influence on molecular-cloud scales, as turbulence and rotation can randomize a cloud's internal field orientation \citep[e.g.,][]{Dobbs2008}. In contrast, other models \citep[e.g.,][]{Shetty2006} propose that a sufficiently strong Galactic field can guide cloud accumulation and fragmentation, imprinting an ordered magnetic structure on the resulting filaments.
Our analysis confirms that the Galactic plane is dominated by a large-scale B-field, as evidenced by the near-parallel alignment of the B-field with the plane (i.e., $\theta_{BG} \sim$0--25$^\circ$). Since the MWLFs are located close to the plane ($|b|<2^\circ$), one might expect the field to influence filament orientation. However, the filament orientations do not strictly follow a plane-aligned trend; instead, they show both parallel and perpendicular cases, suggesting that the large-scale Galactic field is not the sole or dominant factor governing cloud accumulation.
Therefore, the formation of MWLFs might also be influenced by other physical processes.

As discussed in Sec.\,\ref{sec:intro}, a preferential parallel or perpendicular alignment of B-fields with respect to filaments is typically interpreted as evidence for dynamically important B-fields, where the competition between gravitational and turbulent pressures in a magnetically dominated medium shapes the cloud orientation either along or across the field. However, for the MWLFs, we do not observe such preferential alignment; instead, the orientations appear largely random (i.e., $\theta_{FB} \sim$10--80$^\circ$). 
This behavior is consistent with a formation scenario involving shock compression and large-scale gas flows operating in super-Alfvénic conditions, such as those primarily driven by supernova explosions, which may also generate turbulent motions \citep[e.g.,][]{Padoan2004,Hartmann2001,Arzoumanian2011,Inutsuka2015}. 
In such environments, B-fields are dynamically subdominant, allowing gas compression to proceed largely independent of the large-scale inter-cloud B-field, and naturally leading to the absence of a preferred filament–B-field alignment.
In fact, the MWLF sample shows signatures of super-Alfvénic conditions. Based on the Galactic Ring Survey \citep[GRS;][]{Jackson2006} and APEX $^{13}$CO survey data \citep[SEDIGISM;][]{Schuller2017, Schuller2021, Duarte2021} wherever available, we found Alfv\'{e}n Mach numbers of $\mathcal{M}_{\rm A} \gtrsim 2$ in the immediate environments of at least 14 MWLFs. The Alfv\'{e}n Mach number is defined as $\mathcal{M}_{\rm A} = v / v_{\rm A}$, where $v = \sqrt{3}\sigma_v$ is the three-dimensional turbulent velocity dispersion estimated from the observed one-dimensional velocity dispersion $\sigma_v$, and $v_{\rm A} = B / \sqrt{4\pi\rho}$ is the Alfv\'{e}n speed, calculated using the measured B-field strength $B$ and gas density $\rho$. These results are also summarized in Table~\ref{tab:act_filament_results}.
These super-Alfvénic conditions may arise from feedback-driven flows or associated turbulent motions.
Additionally, we find that MWLFs are, on average, offset by $\sim0.5^\circ$ from known Galactic supernova remnants \citep{Green2025}. Overall, the formation of MWLFs is likely linked to large-scale supernova-driven bubble shells. However, a more detailed assessment of the gaseous counterparts is required, which will be investigated in future studies.

We also observe ongoing filament fragmentation in the MWLFs, evident from the presence of dense clumps identified in \textit{Herschel} data. ACT observations further allow a relative angle analysis of the B-field with respect to the filament spine toward 108 clump regions across 20 MWLFs, supporting this fragmentation signature. Figure~\ref{fig:histogram_clump_anlge} shows the histogram distribution of relative angles for clump and filament regions (excluding clumps; see Figure~\ref{fig:f4case} for illustration). The broad distribution ranging from $0^\circ$ to $90^\circ$ for clump regions hints at more random B-field orientations, possibly driven by the collapse of star-forming structures under gravity \citep[e.g.,][]{Sanhueza2021}. 
This is also consistent with studies finding no preferred alignment between outflow axes and both the host filament and local B-field at protostellar scales within evolved protoclusters \citep{Baug2020}.
In contrast, the relative orientation in non-clump regions largely preserves the large-scale filamentary morphology, as seen in Figure~\ref{fig2:relative_bangle_dist_act}. Future high-resolution B-field observations will be valuable for investigating clump-scale B-field patterns.

\section{Conclusions}\label{sec:conc}

We analyzed the relative orientations between MWLFs and the ambient magnetic field using ACT DR6 polarization observations at 220 GHz, complemented by Planck 353 GHz data. 
Overall, the filaments show no strong preferential alignment with the magnetic field. ACT, however, exhibits a modest peak at $50$–$60^\circ$, likely reflecting the smaller number of MWLFs covered by ACT (20) compared to \textit{Planck} (42).
The magnetic field is preferentially aligned with the Galactic plane (e.g., $\theta_{\rm BG}\sim0$–$25^\circ$), whereas filament orientations display a bimodal distribution, with both parallel and perpendicular alignments relative to the plane (e.g., $\theta_{\rm FG}\sim0$–$15^\circ$ and $\sim75$–$90^\circ$). 
We identify two populations of filaments when separated by near and far vertical heights with respect to the Galactic plane.
Filaments located far from the Galactic midplane ($|z|>90$ pc) are preferentially perpendicular to both the plane and the magnetic field, while those near the midplane ($|z|\leq90$ pc) exhibit a mixed population of parallel and perpendicular alignments.
These results suggest that the large-scale Galactic magnetic field does not dominate MWLF formation. The largely random filament–magnetic-field orientations favor a super-Alfvénic formation scenario in which filaments arise from shock compression driven by supernova explosions. This interpretation is consistent with the inferred Alfvénic Mach numbers ($\mathcal{M}_{\rm A} \gtrsim 2$) of MWLFs and their typical angular proximity ($\sim0.5^\circ$) to known Galactic supernova remnants.
Collectively, our findings support a picture in which supernova feedback structures the Galactic disk through a network of bubbles, shaping the large-scale filamentary morphology in a manner reminiscent of nearby face-on spiral galaxies such as M74, as recently revealed by JWST.

\clearpage
\begin{acknowledgements}

We thank the anonymous referee for the helpful comments and suggestions, which have improved the quality of this paper.
We thank Brandon Hensley and Sigurd Naess for the useful suggestions on the usage of ACT data, and Wenyu Jiao, Tapas Baug, Jonathan C. Tan, Dana Alina, and Xing Lu for valuable discussion.
This work has been supported by the National SKA Program of China (2025SKA0140100), National Natural Science Foundation of China (No. 12573025), the Tianshan Talent Training Program (2024TSYCTD0013), China-Chile Joint Research Fund (CCJRF No. 2211), the Tianchi Talent Program of Xinjiang Uygur Autonomous Region, the China Postdoctoral Science Foundation (grant No. 2025M773187), and the High-Performance Computing Platform of Peking University.

\software{Astropy \citep{astropy2013, astropy2018, astropy2022}, Matplotlib \citep{Hunter2007}, NumPy \citep{Harris2020}, SciPy \citep{Virtanen2020}, SAOImageDS9 \citep{Joye2003}}

\end{acknowledgements}

\appendix
\setcounter{figure}{0}
\renewcommand{\thefigure}{A\arabic{figure}}

\section{Background-Corrected filament-B field orientation}
\label{sec:fg_bg_correction}
To confirm whether the relative orientation between filaments and the magnetic field ($\theta_{\rm FB}$) changes significantly after correcting for foreground and background polarization contributions, we derived foreground- and background-corrected Stokes parameter maps following \citet{Alina2019}.
This approach assumes uniform background emission on both sides of the filament and optically thin dust, which is valid for ACT 220\,GHz and \textit{Planck} 353\,GHz.
For a filament region, the observed Stokes parameters at the $i^{\rm th}$ pixel are expressed as
$
X_i = X_{{\rm fil},i} + X_{{\rm bkg},i}, 
$
where  $X = \{I,Q,U\}$ represent the Stokes parameters, and the subscripts `fil' and `bkg' denote the filament and background, respectively. 
For each filament, we defined two background regions on either side, each twice the filament width and offset by $1.5$ widths; we collectively refer to these as the “background (BKG).” 
The BKG region for a reference filament F4 is illustrated in Figure~\ref{fig:f4case}.
BKG-corrected Stokes maps, $X_{\rm corr} = \{I,Q,U\}$, were obtained by subtracting the median BKG Stokes values from the filament-region Stokes maps. The BKG-corrected relative angles ($\theta_{FB,{\rm corr}}$) were then measured from these corrected maps following the approach discussed in Sec.~\ref{sec:obs}.
A comparison of the KDE distributions of the Stokes parameters and derived quantities
(e.g., the polarization fraction,
$p = \sqrt{Q^2 + U^2}\,/\,I$,
$\theta_{B}$, and $\theta_{\rm FB}$)
for the filament, BKG, and BKG-corrected cases of the F4 filament
is presented in Figure~\ref{fig:stokes_dist}.
The corresponding measurements for all MWLFs, along with additional derived properties such as $N_{\rm H_2}$, velocity dispersion (Bhadari et al., in prep), and filament parameters (distance and size from \citealt{Wang2024}), are summarized in Table~\ref{tab:act_filament_results}.

\begin{figure*}[htb!]
	\centering
	\includegraphics[width=1\textwidth]{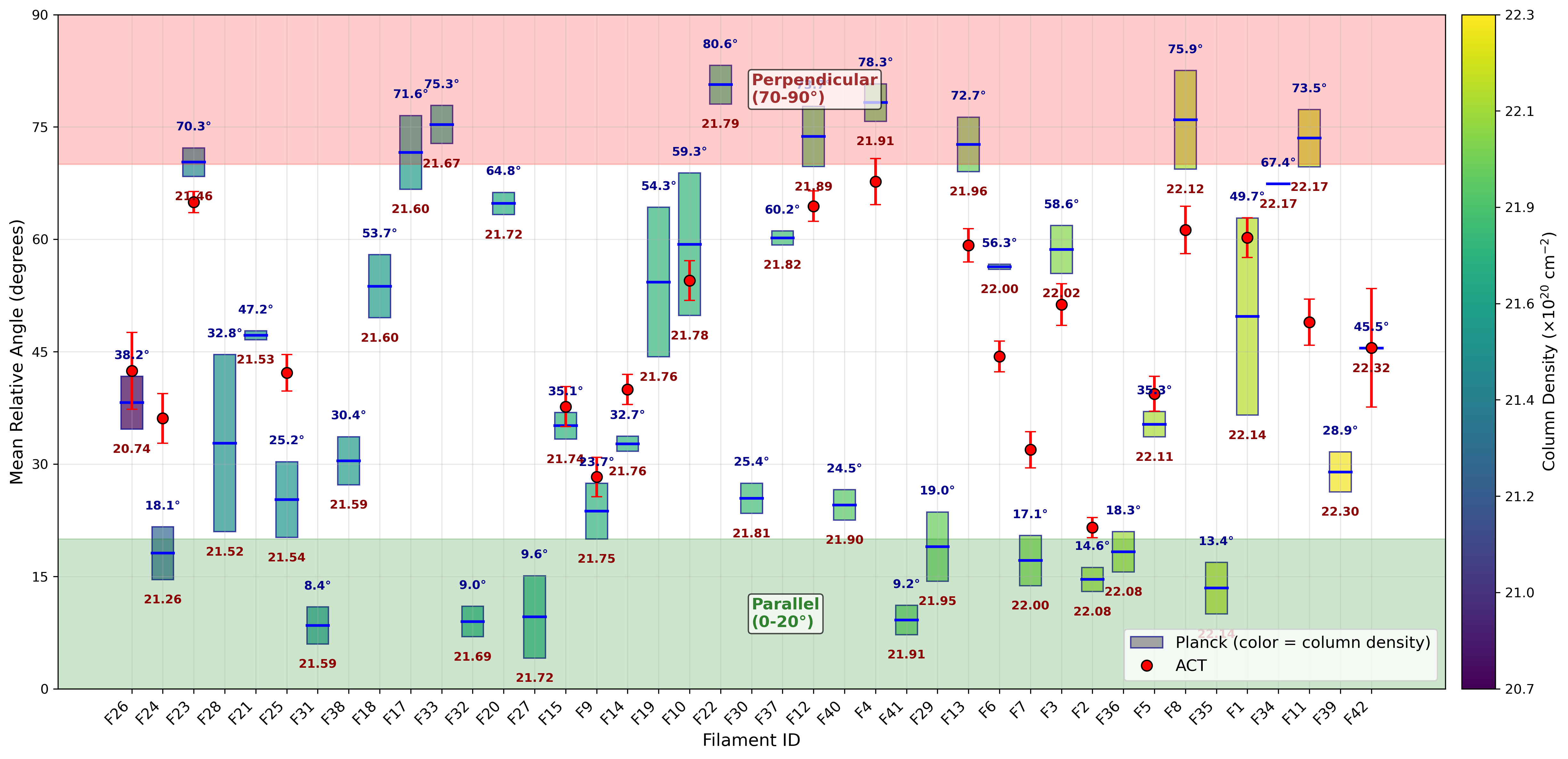}
	\includegraphics[width=1\textwidth]{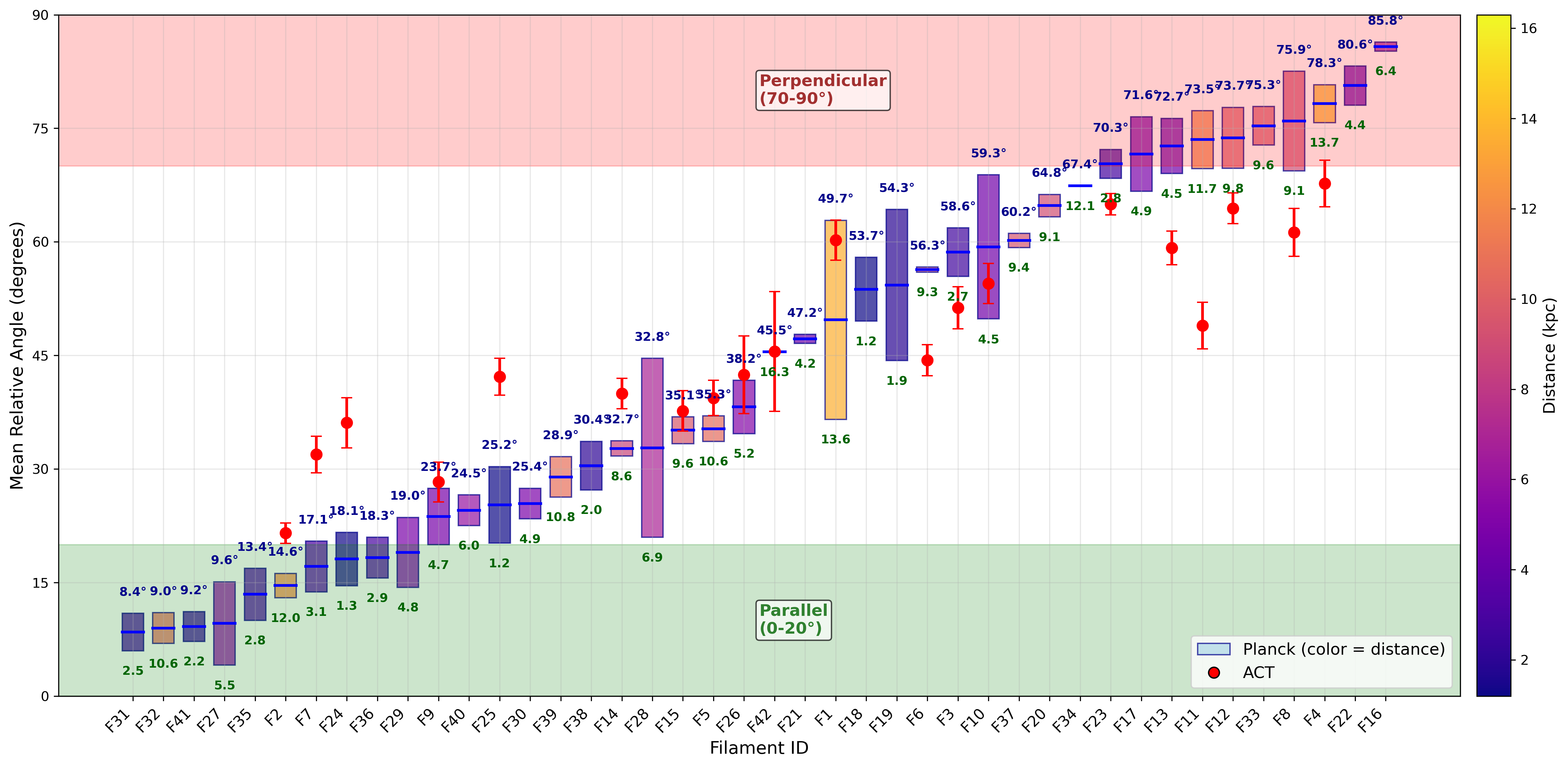}
	\caption{
		Comparison of filament orientations relative to the ambient magnetic field. Colored boxes show the mean relative angles $\pm$ standard error of the mean (SEM) derived from \textit{Planck}, while red circles represent ACT measurements (see Figure~\ref{fig2:relative_bangle_dist_act}). Top labels indicate relative angles in degrees.
		Top panel: Filaments are ordered by increasing H$_2$ column density, with colors and lower-axis labels indicating $N_{\rm H_2}$ in units of 10$^{20}$ cm$^{-2}$.
		Bottom panel: Filaments are ordered by increasing relative angle, with colors and lower labels indicating filament distances in kpc.
	}
	\label{fig:act_planck_relanngle}
\end{figure*}

\begin{figure*}[htb!]
\centering
\includegraphics[width=0.9\textwidth]{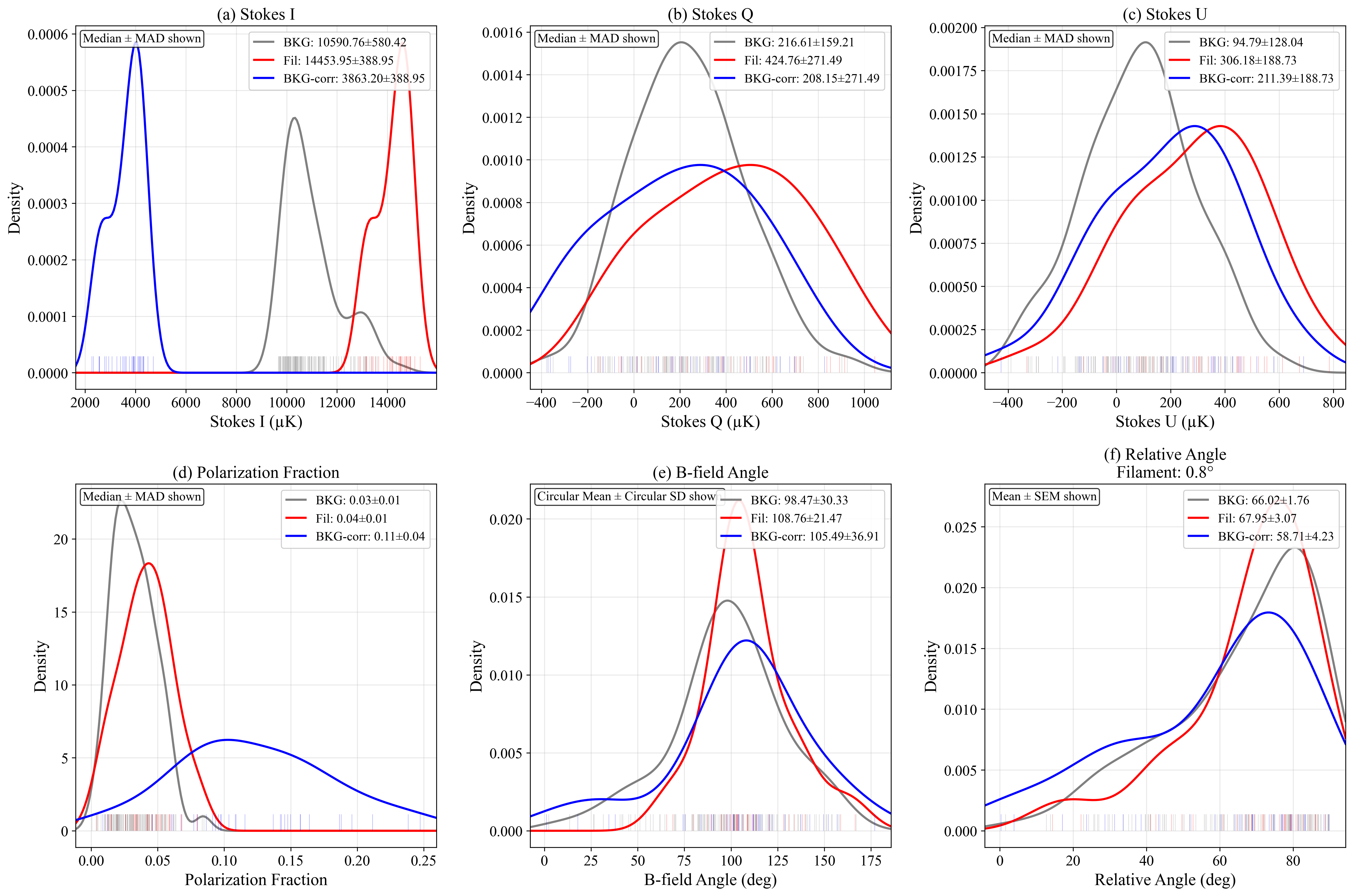}
\caption{
	For filament F4, kernel density distributions of the ACT Stokes parameters $I$, $Q$, and $U$, along with the derived quantities, polarization fraction, POS magnetic field angle, and relative angle between the filament and the magnetic field are shown for the filament (Fil) and background (BKG) subregions, including the background-corrected (BKG-corr) filament case. The median values of each Stokes parameter distribution are indicated, while the circular mean of the B-field angle and the mean relative angle are also labeled.}
\label{fig:stokes_dist}
\end{figure*}

\begin{figure}[htb!]
	\centering
	\includegraphics[width=0.5\linewidth]{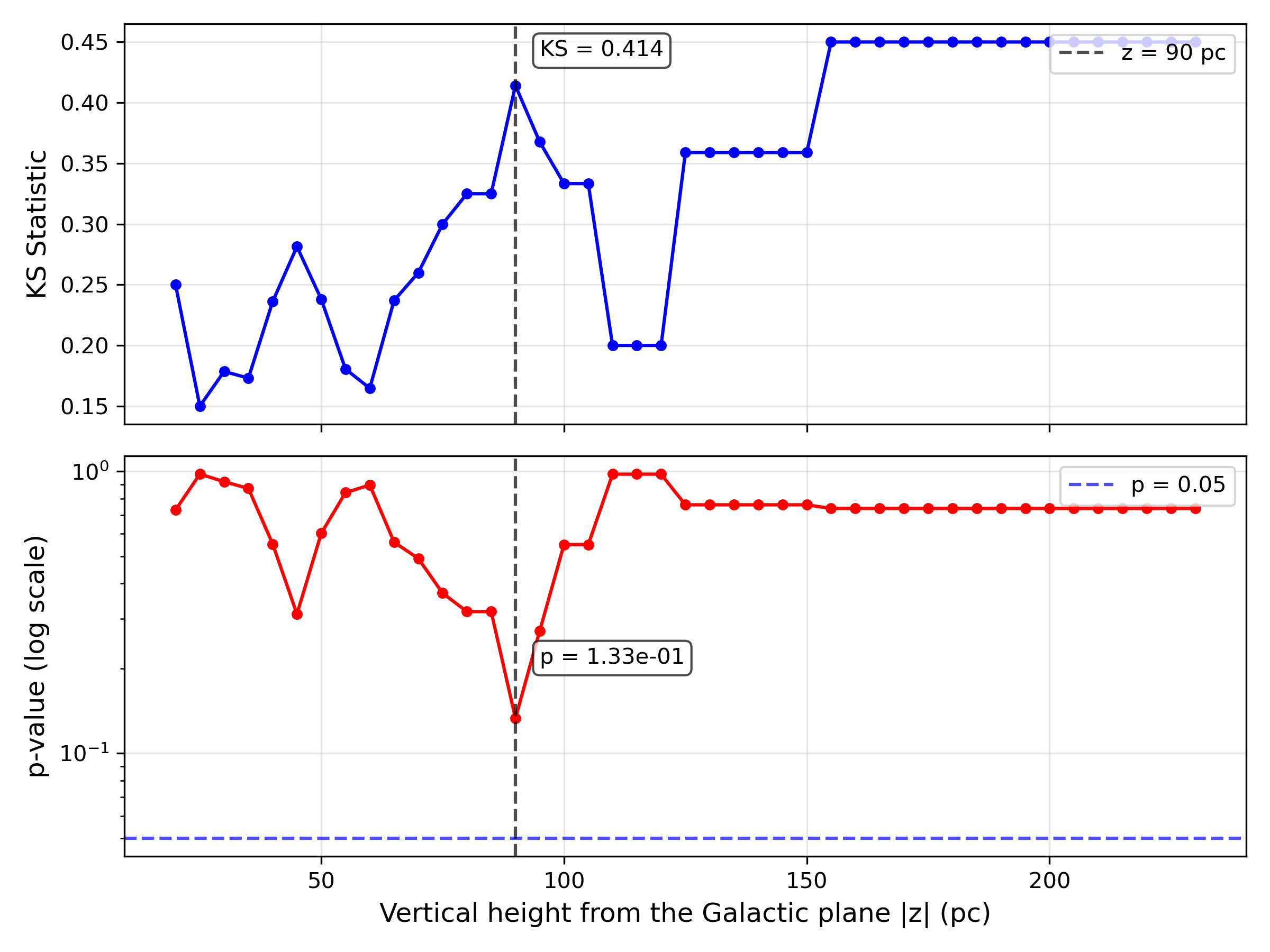}
	\hfill
	\includegraphics[width=0.45\linewidth]{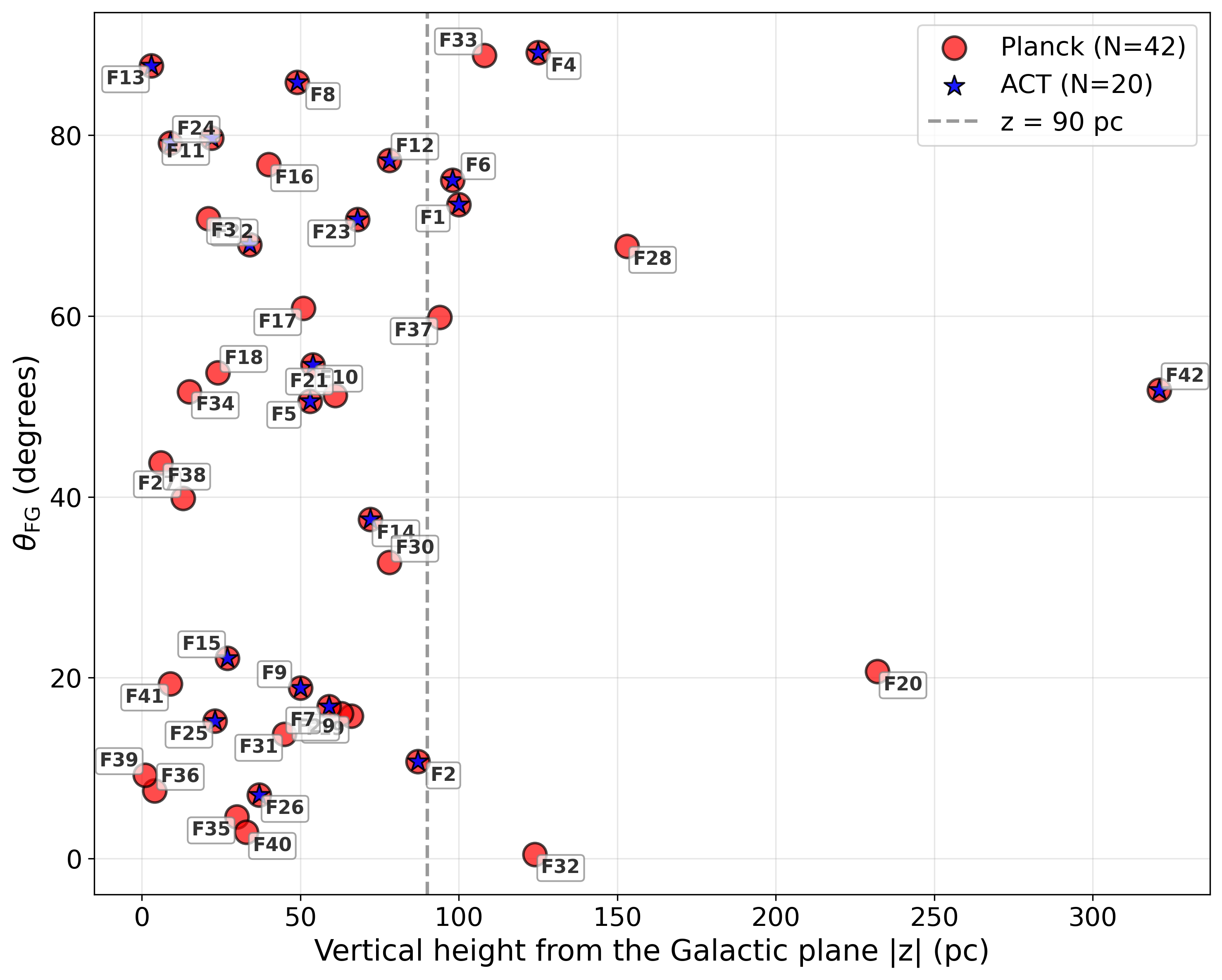}
	\caption{Left: Kolmogorov--Smirnov test statistic (top) and corresponding p-value (bottom) comparing the distributions of $\theta_{\rm FG}$ for filaments below and above each vertical height threshold $|z|$. The vertical dashed line marks $z = 90$ pc, with the KS statistic and p-value annotated. Right: Distribution of $\theta_{\rm FG}$ as a function of vertical height $|z|$.}
	\label{fig:relangle_height}
\end{figure}

\begin{figure}[htb!]
\centering
\includegraphics[width=0.7\linewidth]{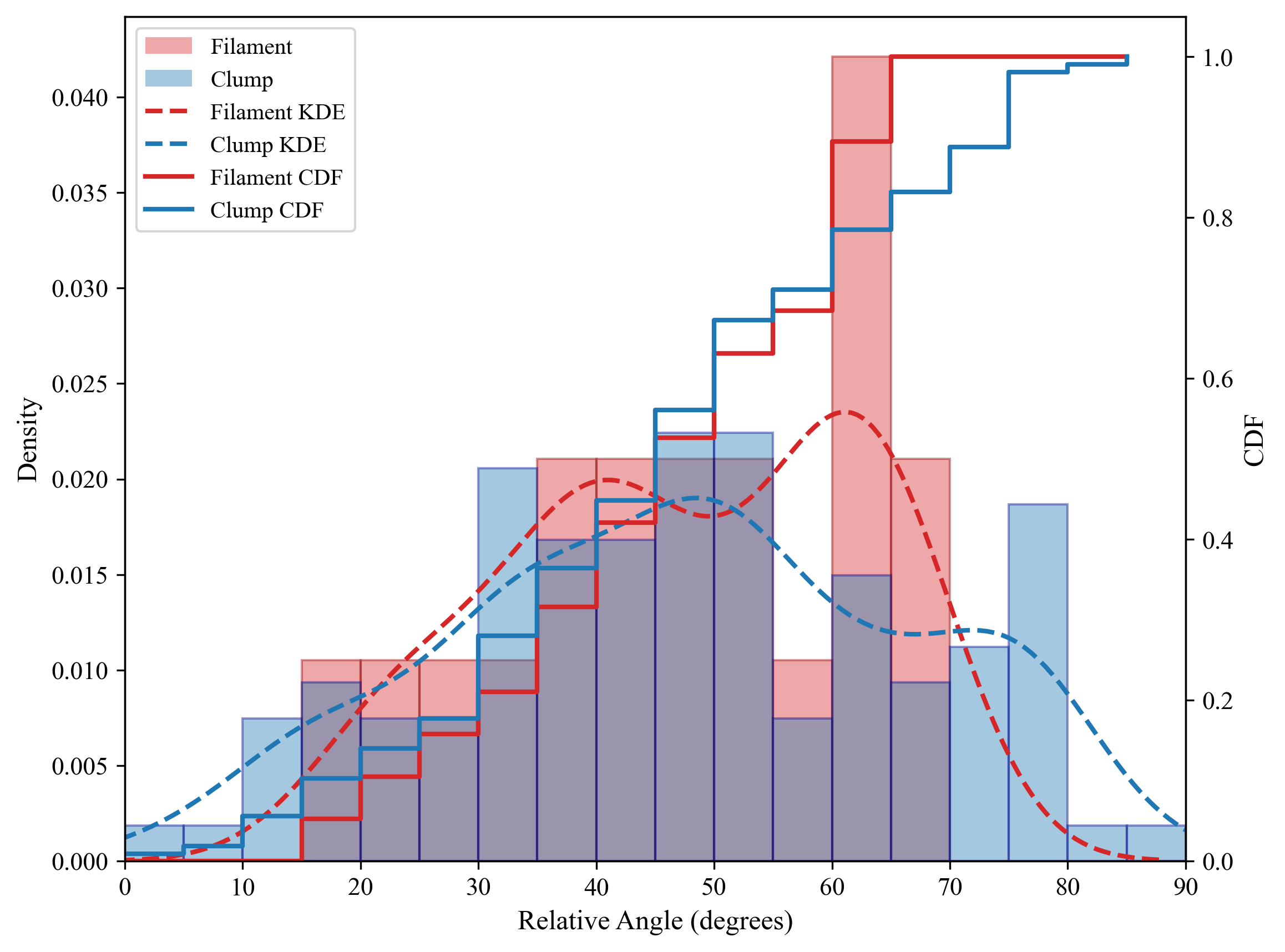}
\caption{Distribution of B-field relative angles in filament and clump regions. Histograms (bars) and kernel density estimates (dashed curves) show the orientation distributions, while stepwise cumulative distribution functions (solid curves) show the cumulative probabilities. Filament regions are shown in red, clump regions in blue ($n = 108$ clumps across $20$ filaments).
}
\label{fig:histogram_clump_anlge}
\end{figure}

\begin{figure}[htb!]
	\centering
	\includegraphics[width=1\linewidth]{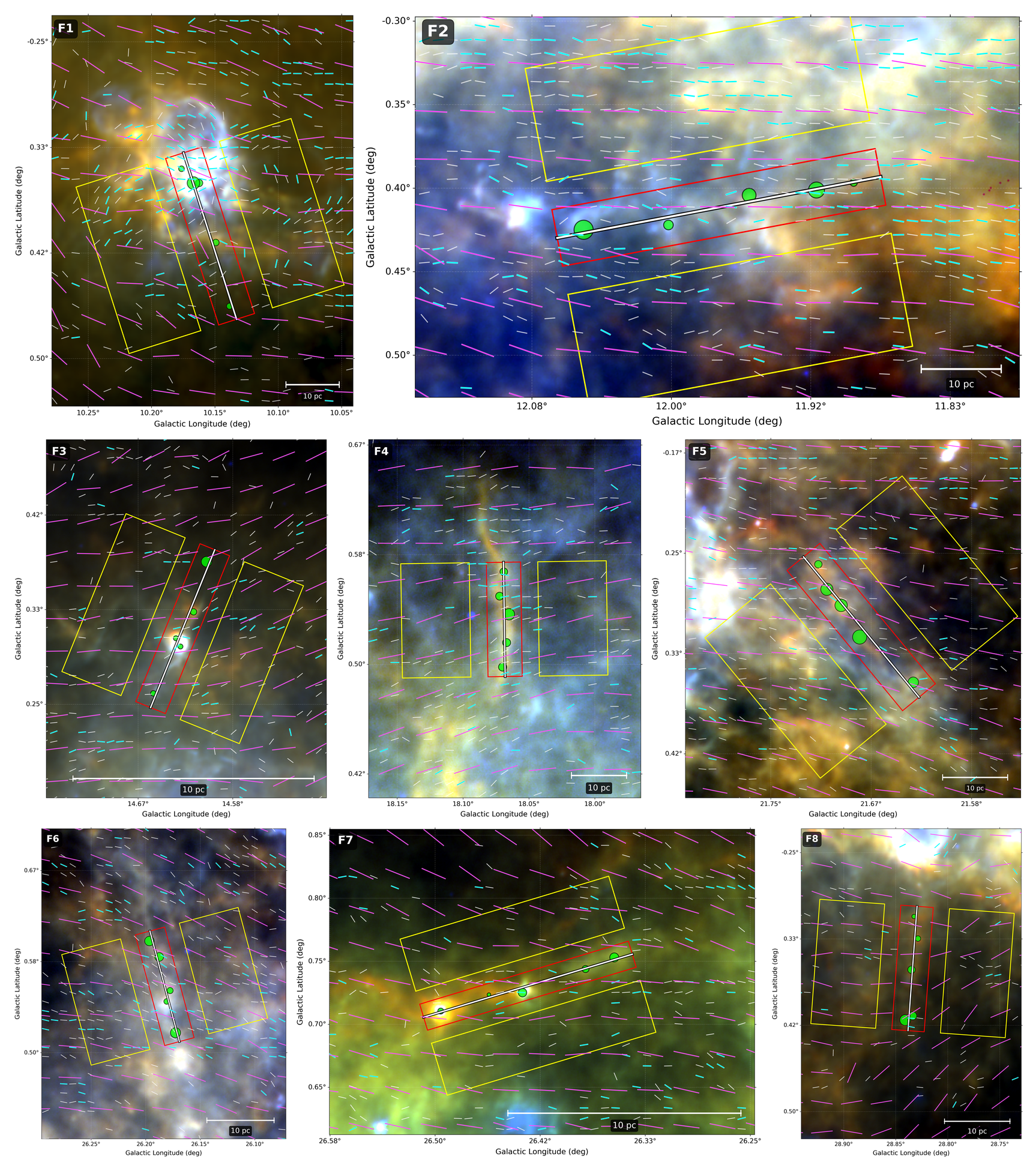}
	\caption{
		Plane-of-sky magnetic-field morphology of filaments F1--F8 as traced by ACT observations at 220\,GHz (white: $1.5 \leq P/\sigma_{\rm P} < 2$; cyan: $P/\sigma_{\rm P} \geq 2$). 
		The background image is an RGB composite constructed from \textit{Herschel} 250 (Red), 160 (Green), and 70~$\mu$m (Blue) emission. 
		The overlaid elements are the same as those shown in Figure~\ref{fig:f4case}, with the addition of magnetic-field vectors derived from \textit{Planck} 353\,GHz polarization data, shown in magenta. A two-color (\textit{Herschel} 250~$\mu$m [red] and 70~$\mu$m [cyan]) view of all the studied targets can be found in \citet{Wang2024}.
}
	\label{fig:collage1}
\end{figure}

\begin{figure}[htb!]
	\centering
	\includegraphics[width=1\linewidth]{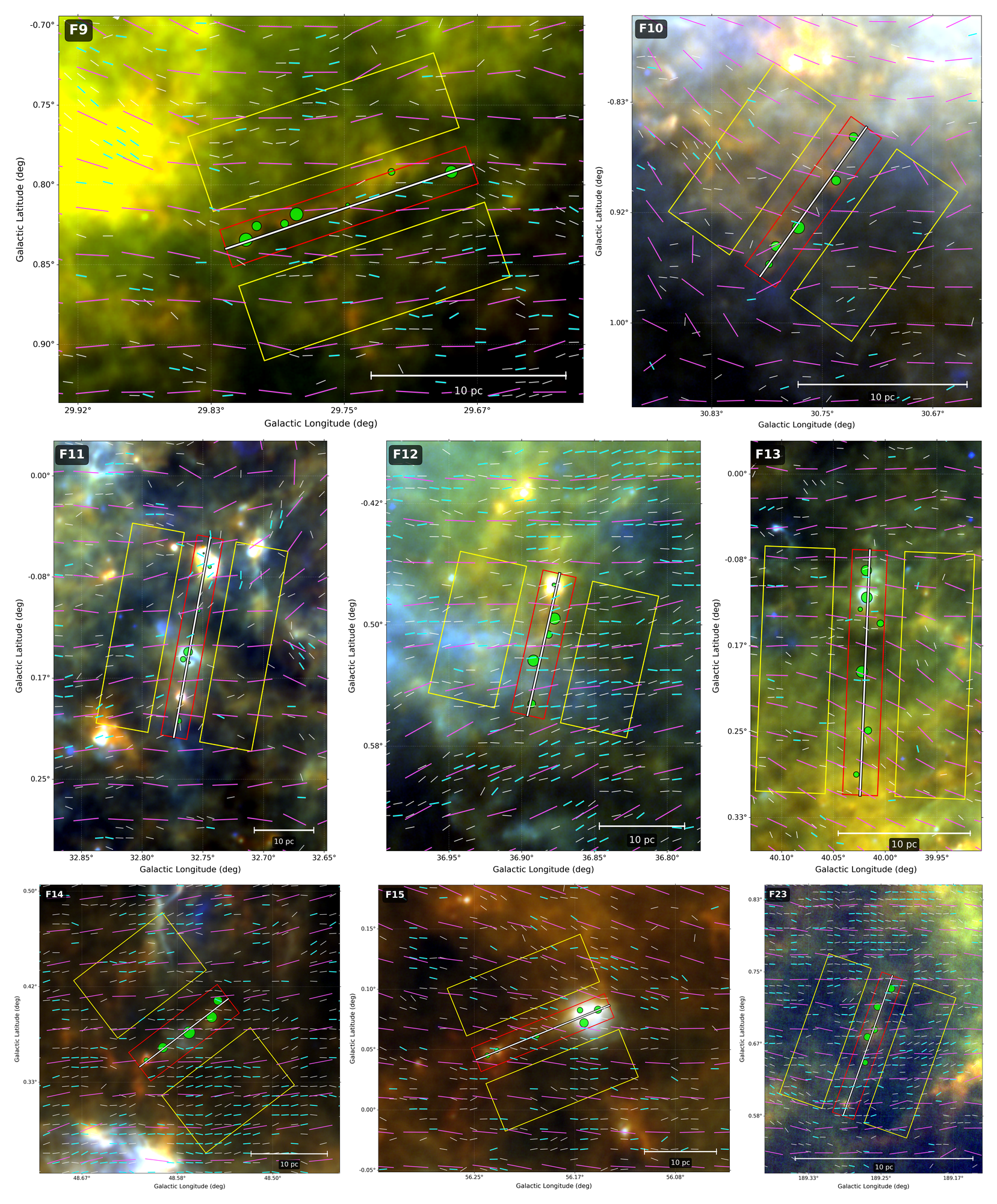}
	\caption{
		Same as Figure~\ref{fig:collage1}, but for filaments F9--F15 and F23.
	}
	\label{fig:collage2}
\end{figure}

\begin{figure}[htb!]
	\centering
	\includegraphics[width=1\linewidth]{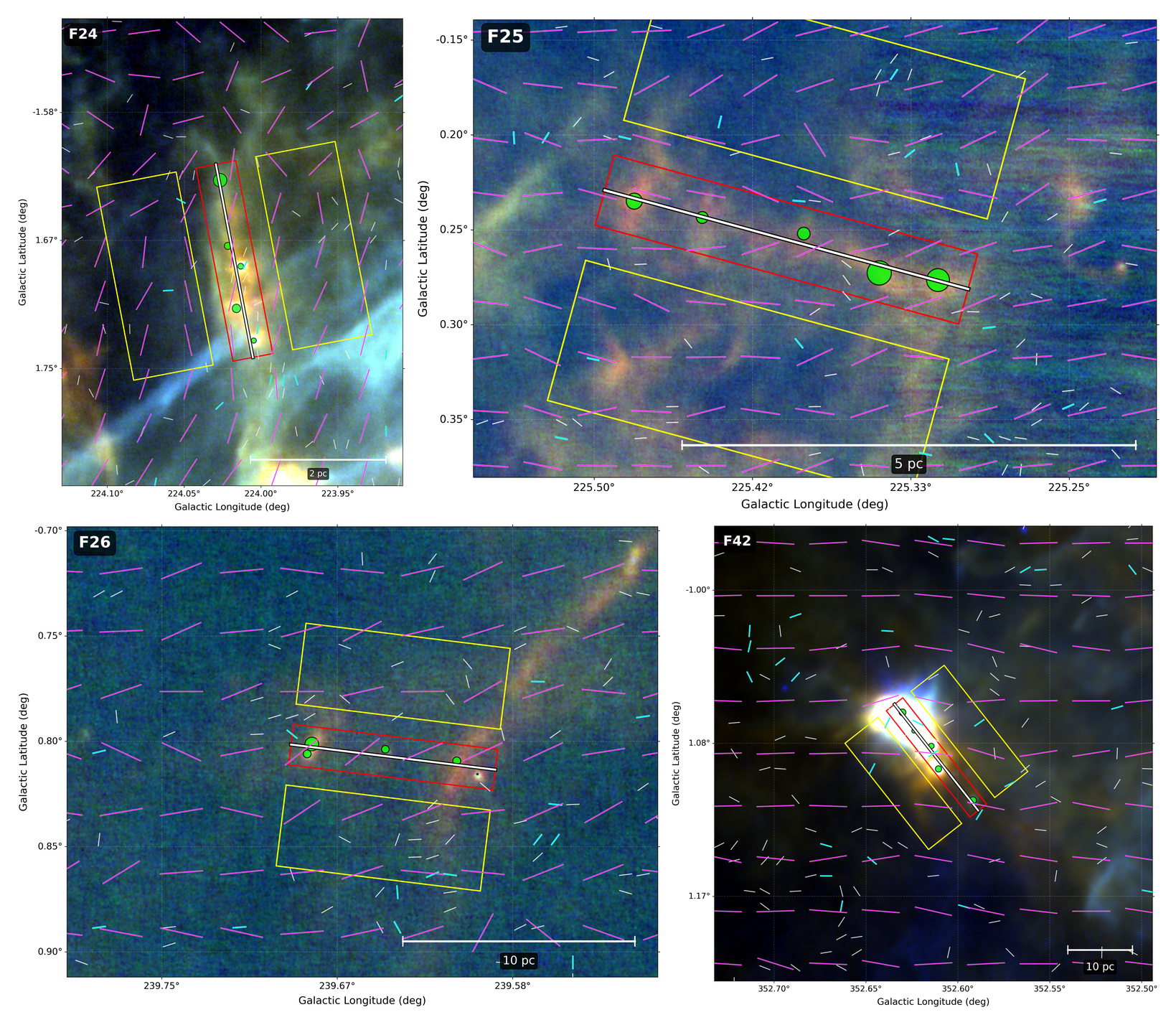}
	\caption{
			Same as Figure~\ref{fig:collage1}, but for filaments  F24, F25, F26, and F42.
	}
	\label{fig:collage3}
\end{figure}

\begin{sidewaystable}
	\centering
	\scriptsize
	\setlength{\tabcolsep}{0.5pt}  
	\caption{Derived Physical Parameters of Giant Filaments Traced by ACT}
	\label{tab:act_filament_results}
	\begin{tabular}{lcccccccccccccccccccccccccc}
		\toprule
		
		Fil. & $l$ & $b$ & $d$ & $L$ & $z$ & $\theta_F$ & $I_f$ & $Q_f$ & $U_f$ & $p_f$ & $\theta_{\mathrm{B},f}$ & $\theta_{\mathrm{FB},f}$ & $I_b$ & $Q_b$ & $U_b$ & $p_b$ & $\theta_{\mathrm{B},b}$ & $\theta_{\mathrm{FB},b}$ & $\theta_{\mathrm{FB,corr}}$ & $\theta_{\mathrm{FG}}$ & $\theta_{\mathrm{BG}}$ & $N_{\mathrm{H_2},f}$ & $\sigma_{v,f}$ & $\mathcal{M}_{\mathrm{A},f}$ & $\mathcal{M}_{\mathrm{A},b}$ \\
		ID & ($^\circ$) & ($^\circ$) & (kpc) & (pc) & (pc) & ($^\circ$) & ($\mu$K) & ($\mu$K) & ($\mu$K) & & ($^\circ$) & ($^\circ$) & ($\mu$K) & ($\mu$K) & ($\mu$K) & & ($^\circ$) & ($^\circ$) & ($^\circ$) & ($^\circ$) & ($^\circ$) & ($\times 10^{20}$ cm$^{-2}$) & (km/s) & & \\
		
		\midrule
		
		F1 & 10.2$^\circ$ & -0.4$^\circ$ & 13.6 & 33.0 & -100.0 & 17.7$^\circ$ & $4.55 \times 10^{4}$ & $215.92$ & $130.70$ & $0.010$ & $105.1^\circ$ & $60.2^\circ$ & $3.11 \times 10^{4}$ & $146.63$ & $118.95$ & $0.012$ & $101.7^\circ$ & $59.2^\circ$ & $51.7^\circ$ & $72.3^\circ$ & $15.6^\circ$ & $22.14$ & $5.68$ & 2.12 & 1.92 \\
		F2 & 12.0$^\circ$ & -0.4$^\circ$ & 12.0 & 39.5 & -87.0 & 100.7$^\circ$ & $1.65 \times 10^{4}$ & $496.42$ & $-38.48$ & $0.034$ & $88.1^\circ$ & $21.5^\circ$ & $1.65 \times 10^{4}$ & $448.59$ & $-39.74$ & $0.030$ & $87.9^\circ$ & $17.9^\circ$ & $41.8^\circ$ & $10.7^\circ$ & $2.2^\circ$ & $22.08$ & $0.20$ & 1.21 & 1.53 \\
		F3 & 14.6$^\circ$ & 0.4$^\circ$ & 2.7 & 7.0 & 34.0 & 157.9$^\circ$ & $1.75 \times 10^{4}$ & $228.01$ & $89.14$ & $0.022$ & $95.5^\circ$ & $51.3^\circ$ & $1.28 \times 10^{4}$ & $201.06$ & $146.03$ & $0.028$ & $106.0^\circ$ & $46.9^\circ$ & $49.0^\circ$ & $67.9^\circ$ & $10.7^\circ$ & $22.02$ & $0.65$ & 2.11 & 1.94 \\
		F4 & 18.1$^\circ$ & 0.6$^\circ$ & 13.7 & 22.0 & 125.0 & 0.8$^\circ$ & $1.45 \times 10^{4}$ & $424.76$ & $306.18$ & $0.042$ & $108.8^\circ$ & $67.7^\circ$ & $1.06 \times 10^{4}$ & $216.61$ & $94.79$ & $0.031$ & $98.5^\circ$ & $66.0^\circ$ & $58.5^\circ$ & $89.2^\circ$ & $17.9^\circ$ & $21.91$ & $0.32$ & 1.26 & 1.94 \\
		F5 & 21.7$^\circ$ & -0.3$^\circ$ & 10.6 & 28.1 & -53.0 & 39.4$^\circ$ & $2.05 \times 10^{4}$ & $264.17$ & $-141.92$ & $0.019$ & $77.4^\circ$ & $39.4^\circ$ & $1.89 \times 10^{4}$ & $324.06$ & $-117.81$ & $0.021$ & $80.0^\circ$ & $40.9^\circ$ & $42.6^\circ$ & $50.6^\circ$ & $14.1^\circ$ & $22.11$ & $0.49$ & 2.54 & 1.85 \\
		F6 & 26.2$^\circ$ & 0.6$^\circ$ & 9.3 & 17.6 & 98.0 & 15.0$^\circ$ & $1.47 \times 10^{4}$ & $204.46$ & $-324.29$ & $0.030$ & $58.6^\circ$ & $44.4^\circ$ & $1.32 \times 10^{4}$ & $304.53$ & $-178.20$ & $0.032$ & $72.2^\circ$ & $55.5^\circ$ & $30.3^\circ$ & $75.0^\circ$ & $28.9^\circ$ & $22.00$ & $0.32$ & 3.83 & 2.88 \\
		F7 & 26.5$^\circ$ & 0.7$^\circ$ & 3.1 & 8.6 & 59.0 & 106.8$^\circ$ & $1.60 \times 10^{4}$ & $222.69$ & $-45.30$ & $0.018$ & $84.4^\circ$ & $31.9^\circ$ & $1.11 \times 10^{4}$ & $199.53$ & $-78.23$ & $0.028$ & $79.5^\circ$ & $32.2^\circ$ & $40.2^\circ$ & $16.8^\circ$ & $5.7^\circ$ & $22.00$ & $0.70$ & 2.40 & 2.64 \\
		F8 & 28.8$^\circ$ & -0.3$^\circ$ & 9.1 & 19.9 & -49.0 & 175.9$^\circ$ & $1.97 \times 10^{4}$ & $191.39$ & $-57.54$ & $0.014$ & $85.3^\circ$ & $61.2^\circ$ & $1.72 \times 10^{4}$ & $143.15$ & $-11.42$ & $0.016$ & $88.9^\circ$ & $61.1^\circ$ & $49.4^\circ$ & $85.9^\circ$ & $8.4^\circ$ & $22.12$ & $0.50$ & 2.76 & 2.74 \\
		F9 & 29.8$^\circ$ & -0.8$^\circ$ & 4.7 & 13.1 & -50.0 & 108.9$^\circ$ & $1.10 \times 10^{4}$ & $201.55$ & $29.59$ & $0.026$ & $92.7^\circ$ & $28.3^\circ$ & $1.07 \times 10^{4}$ & $199.80$ & $34.02$ & $0.023$ & $92.7^\circ$ & $23.3^\circ$ & $46.2^\circ$ & $18.9^\circ$ & $4.2^\circ$ & $21.75$ & $0.32$ & 2.12 & 1.79 \\
		F10 & 30.7$^\circ$ & -0.9$^\circ$ & 4.5 & 10.8 & -54.0 & 144.6$^\circ$ & $1.01 \times 10^{4}$ & $89.54$ & $-68.21$ & $0.024$ & $69.7^\circ$ & $54.5^\circ$ & $8.33 \times 10^{3}$ & $104.46$ & $-22.47$ & $0.024$ & $84.2^\circ$ & $51.9^\circ$ & $48.4^\circ$ & $54.6^\circ$ & $18.6^\circ$ & $21.78$ & $0.32$ & 2.83 & 2.60 \\
		F11 & 32.7$^\circ$ & -0.1$^\circ$ & 11.7 & 32.8 & -22.0 & 169.7$^\circ$ & $2.22 \times 10^{4}$ & $129.07$ & $172.07$ & $0.013$ & $112.3^\circ$ & $48.9^\circ$ & $1.83 \times 10^{4}$ & $83.49$ & $61.72$ & $0.012$ & $101.2^\circ$ & $50.3^\circ$ & $43.3^\circ$ & $79.7^\circ$ & $26.6^\circ$ & $22.17$ & $0.69$ & 2.07 & 2.12 \\
		F12 & 36.9$^\circ$ & -0.5$^\circ$ & 9.8 & 17.2 & -78.0 & 167.2$^\circ$ & $1.90 \times 10^{4}$ & $351.92$ & $128.84$ & $0.022$ & $101.5^\circ$ & $64.4^\circ$ & $1.47 \times 10^{4}$ & $352.16$ & $76.47$ & $0.026$ & $95.0^\circ$ & $69.3^\circ$ & $43.8^\circ$ & $77.2^\circ$ & $10.1^\circ$ & $21.89$ & $0.52$ & 0.85 & 0.99 \\
		F13 & 40.0$^\circ$ & -0.1$^\circ$ & 4.5 & 18.6 & 3.0 & 177.7$^\circ$ & $1.75 \times 10^{4}$ & $162.74$ & $-96.80$ & $0.014$ & $69.6^\circ$ & $59.2^\circ$ & $1.68 \times 10^{4}$ & $145.32$ & $-46.72$ & $0.015$ & $81.2^\circ$ & $61.0^\circ$ & $45.3^\circ$ & $87.7^\circ$ & $15.4^\circ$ & $21.96$ & $0.52$ & 2.80 & 2.68 \\
		F14 & 48.6$^\circ$ & 0.4$^\circ$ & 8.6 & 15.7 & 72.0 & 127.5$^\circ$ & $1.08 \times 10^{4}$ & $312.20$ & $-10.07$ & $0.031$ & $89.1^\circ$ & $39.9^\circ$ & $1.03 \times 10^{4}$ & $293.85$ & $87.32$ & $0.033$ & $98.2^\circ$ & $29.9^\circ$ & $59.0^\circ$ & $37.5^\circ$ & $0.9^\circ$ & $21.76$ & $0.32$ & 0.91 & 1.30 \\
		F15 & 56.2$^\circ$ & 0.1$^\circ$ & 9.6 & 21.3 & 27.0 & 112.2$^\circ$ & $1.02 \times 10^{4}$ & $274.00$ & $-131.42$ & $0.031$ & $74.4^\circ$ & $37.7^\circ$ & $7.00 \times 10^{3}$ & $239.99$ & $-142.56$ & $0.045$ & $78.7^\circ$ & $36.4^\circ$ & $42.6^\circ$ & $22.2^\circ$ & $12.8^\circ$ & $21.74$ & -- & -- & -- \\
		F23 & 189.2$^\circ$ & 0.7$^\circ$ & 2.8 & 8.0 & 68.0 & 160.7$^\circ$ & $3.94 \times 10^{3}$ & $204.12$ & $-4.43$ & $0.054$ & $90.1^\circ$ & $65.0^\circ$ & $3.34 \times 10^{3}$ & $202.74$ & $8.82$ & $0.068$ & $91.0^\circ$ & $66.2^\circ$ & $46.4^\circ$ & $70.7^\circ$ & $0.6^\circ$ & $21.46$ & -- & -- & -- \\
		F24 & 224.0$^\circ$ & -1.6$^\circ$ & 1.3 & 2.8 & -9.0 & 10.9$^\circ$ & $6.00 \times 10^{3}$ & $-193.42$ & $76.17$ & $0.065$ & $106.0^\circ$ & $36.1^\circ$ & $3.34 \times 10^{3}$ & $-30.39$ & $34.25$ & $0.093$ & $93.3^\circ$ & $42.6^\circ$ & $36.7^\circ$ & $79.1^\circ$ & $79.3^\circ$ & $21.26$ & -- & -- & -- \\
		F25 & 225.5$^\circ$ & -0.2$^\circ$ & 1.2 & 4.0 & 23.0 & 74.8$^\circ$ & $4.43 \times 10^{3}$ & $-16.52$ & $-25.04$ & $0.072$ & $88.0^\circ$ & $42.2^\circ$ & $3.36 \times 10^{3}$ & $51.87$ & $8.71$ & $0.091$ & $91.1^\circ$ & $41.9^\circ$ & $45.0^\circ$ & $15.2^\circ$ & $61.7^\circ$ & $21.54$ & -- & -- & -- \\
		F26 & 239.7$^\circ$ & -0.8$^\circ$ & 5.2 & 9.1 & -37.0 & 83.0$^\circ$ & $2.33 \times 10^{3}$ & $17.46$ & $-117.18$ & $0.107$ & $71.9^\circ$ & $42.4^\circ$ & $1.52 \times 10^{3}$ & $82.35$ & $40.05$ & $0.192$ & $95.7^\circ$ & $37.0^\circ$ & $48.5^\circ$ & $7.0^\circ$ & $40.8^\circ$ & $20.74$ & -- & -- & -- \\
		F42 & 352.6$^\circ$ & -1.1$^\circ$ & 16.3 & 20.0 & -321.0 & 38.2$^\circ$ & $5.40 \times 10^{4}$ & $478.82$ & $-408.99$ & $0.023$ & $85.5^\circ$ & $45.5^\circ$ & $2.32 \times 10^{4}$ & $253.07$ & $349.82$ & $0.043$ & $105.9^\circ$ & $58.0^\circ$ & $35.8^\circ$ & $51.8^\circ$ & $20.3^\circ$ & $22.32$ & -- & -- & -- \\
		
		\bottomrule
		\end{tabular}
		\vspace{0.2in}
		\begin{flushleft}
		\footnotesize
		\textit{Note:} $l$ and $b$ are Galactic coordinates in degrees. $d$ is distance in kpc. $L$ is filament length in pc. $z$ is height above Galactic plane in pc. $\theta_F$ is filament orientation angle measured from North to East counterclockwise. $I$, $Q$, $U$ are Stokes parameters in $\mu$K. $p$ is polarization fraction. $\theta_{\rm B}$ is B-field angle in degrees. 
		$\theta_{\rm FB}$ and $\theta_{\mathrm{FB,corr}}$ are the relative angles between the filament and the B-field for the standard and background-corrected cases, respectively. These values are derived using the Monte Carlo approach (see Sec.~\ref{sec:relangle_bfield}). $\theta_{\rm FG}$ and $\theta_{\rm BG}$ are the Galactic plane-filament, and Galactic plane-B-field relative angles, respectively. $\sigma_v$ is velocity dispersion in km/s, derived using $^{13}$CO emission from existing molecular line surveys. $N_{\mathrm{H_2}}$ is H$_2$ column density in units of 10$^{20}$ cm$^{-2}$. $\mathcal{M}_{\mathrm{A}}$ represents Alfvén Mach number. Subscripts $f$ and $b$ denote filament and background regions respectively. Stokes parameters show median values, B-field angles show circular mean, relative angles, $N_{\mathrm{H_2}}$, $\sigma_v$, and $\mathcal{M}_{\mathrm{A}}$ show mean.
		\end{flushleft}
	\end{sidewaystable}

\bibliography{references}{}
\bibliographystyle{aasjournal}

\end{document}